%% file: PRL_IPD_revised.tex
\documentclass[prx, aps,amsmath,amssymb,twocolumn,nofootinbib,floatfix]{revtex4-2}
\usepackage[utf8]{inputenc}        
\usepackage{natbib}   
\usepackage{amsmath}
\usepackage{amssymb}
\usepackage{slashed}
\usepackage{adjustbox}
\usepackage{booktabs}
\usepackage[caption=false]{subfig}
\labelformat{subfigure}{\thefigure(#1)}
\usepackage{enumitem} 
\usepackage{makecell}

\def\sint{{\textstyle\sum}\!\!\!\!\!\!\!\int}

\usepackage[colorlinks,linkcolor=blue,anchorcolor=blue,citecolor=blue,urlcolor=blue]{hyperref}

\begin{document}



\title{Towards Consistent Thermal Resummation: Improved Partial Dressing for Precise Phase Transition Predictions}


\author{Dayun Qiu}
\email{qiudy3@mail2.sysu.edu.cn}
\author{Fa Peng Huang}
\email{Corresponding author.  huangfp8@sysu.edu.cn}

\affiliation{MOE Key Laboratory of TianQin Mission, TianQin Research Center for Gravitational Physics \& School of Physics and Astronomy, Frontiers Science Center for TianQin, Gravitational Wave Research Center of CNSA, Sun Yat-sen University (Zhuhai Campus), Zhuhai 519082, China}



\begin{abstract}
Reliable predictions for the phase transition depend critically on a consistent resummation of thermal fluctuations. 
However, existing prescriptions often entail sizable uncertainties, motivating the development of a robust and systematically controlled resummation scheme.
In this letter, we develop an improved partial dressing (IPD) scheme that resolves intrinsic inconsistencies in the original partial dressing  prescription, including the overlapping momentum problem and the associated divergences in the gap equations. By consistently incorporating momentum-dependent self-energies and reorganizing the daisy resummation, the IPD scheme provides a more consistent treatment of thermal resummation. We apply IPD in the well-motivated inert doublet model and compare its predictions with those of the conventional resummation prescriptions. 
These results demonstrate that a self-consistent treatment of thermal fluctuations significantly affects the predicted phase transition parameters and gravitational wave spectra, underscoring the need to control thermal-resummation uncertainties for reliable cosmological predictions.
\end{abstract}

\maketitle

{\it \textbf{Introduction.}} Cosmological first-order phase transition (FOPT) 
might play important roles in many fundamental problems of our early Universe, such as the origin of baryon asymmetry via electroweak baryogenesis~\cite{Cohen:1993nk,Trodden:1998ym,Morrissey:2012db,Cohen:1994ss,Konstandin:2013caa,Wagner:2023vqw,vandeVis:2025efm, Roy:2025zvo}, the production of heavy dark matter~\cite{Baker:2019ndr,Azatov:2021ifm,Chway:2019kft,Giudice:2024tcp,Baldes:2023cih,Chowdhury:2011ga,Huang:2017kzu, Jiang:2023qbm, Jiang:2023nkj,Jiang:2024zrb, Jiang:2025xln}, the production of primordial black holes~\cite{Jedamzik:1999am,Liu:2021svg,Lewicki:2023ioy,Huang:2022him,Goncalves:2024vkj,Hashino:2025fse,Jung:2021mku,Baker:2021sno}, and the generation of gravitational wave (GW) background~\cite{Kamionkowski:1993fg, Caprini:2015zlo,Caprini:2018mtu,Caprini:2019egz,Hindmarsh:2013xza,Hindmarsh:2015qta,Hindmarsh:2017gnf,Cutting:2018tjt,Jinno:2022fom,Qiu:2025tmn,Ai:2025fqw} detectable by LISA~\cite{LISA:2017pwj}, TianQin~\cite{TianQin:2015yph}, Taiji~\cite{Hu:2017mde}, BBO~\cite{Crowder:2005nr,Corbin:2005ny}, Cosmic Explorer (CE)~\cite{Reitze:2019iox}, Ultimate (U)-DECIGO~\cite{Kudoh:2005as}, Einstein Telescope (ET)~\cite{Sathyaprakash:2012jk,ET:2019dnz}, LIGO-Virgo-KAGRA~\cite{LIGOScientific:2014pky,VIRGO:2014yos,KAGRA:2020tym}, and NANOGrav/EPTA/PPTA/CPTA~\cite{NANOGrav:2023gor,EPTA:2023fyk,Reardon:2023gzh,Xu:2023wog}. Reliable predictions for these phenomena require precise finite-temperature effective potential. Its perturbative expansion, however, is infrared (IR) sensitive because of bosonic Matsubara zero modes, making thermal resummation essential.

Various thermal resummation prescriptions and their optimizations have been developed to regulate IR divergences in the finite-temperature effective potential~\cite{Dolan:1973qd, Parwani:1991gq, Arnold:1992rz, Boyd:1993tz, Espinosa:1992gq,Espinosa:1992kf, Amelino-Camelia:1992qfe, Boyd:1992xn, Buchmuller:1993bq, Curtin:2016urg, Bahl:2024ykv, Banik:2026ljf, Farakos:1994kx,Kajantie:1995dw,Braaten:1995cm,Gould:2026rqz, Berthiaume:2026dkt}, differing in the range of diagrams resummed and the set of underlying approximations. Those commonly adopted resummation prescriptions based on the high-temperature expansion are subject to the limitations of this approximation, which can become unreliable at the phase transition temperature.
The partial dressing (PD) prescription~\cite{Boyd:1992xn} provides an alternative that does not rely on the high-temperature approximation. It determines thermal masses through gap equations and reconstructs the effective potential by substituting the resummed masses at the level of its first derivative, thereby avoiding the double counting associated with a global mass replacement. However, the original PD formulation suffers from intrinsic inconsistencies: it neglects the external momentum dependence of the resummed self-energy, an approximation that is justified only for zero modes and requires a specific hierarchy among the loop momenta in the superdaisy diagrams; moreover, the gap equation exhibits a divergent behavior in the zero-mass limit, compromising the smoothness of the effective potential in the vicinity of the tunneling barrier.
Together, these limitations introduce uncontrolled theoretical uncertainties~\cite{Athron:2022jyi,Bittar:2025lcr,Christiansen:2025xhv,Cho:2026pfw}, especially for weak FOPTs lying close to the second-order critical boundary.

In this {\it letter}, we propose an improved partial dressing (IPD) resummation scheme that resolves the aforementioned inconsistencies in the original PD prescription. By restructuring the resummation framework for daisy diagrams and properly accounting for the momentum dependence of the self-energy, the IPD scheme yields a smooth effective potential and more robust phase transition parameters. To illustrate the implementation and demonstrate the practical advantages of the proposed framework, we apply the IPD scheme to the well-motivated inert doublet model (IDM)~\cite{Deshpande:1977rw,Barbieri:2006dq}. We also perform a systematic comparison with several conventional Arnold--Espinosa (AE)~\cite{Arnold:1992rz}, Parwani~\cite{Parwani:1991gq}, dimensional reduction (DR)~\cite{Farakos:1994kx,Kajantie:1995dw,Braaten:1995cm} prescriptions relying on the high-temperature expansion, and the original PD prescription to quantify the impact of different resummation schemes on the predicted phase transition parameters and associated GW signals. Our results reveal substantial scheme dependence, particularly for weak FOPTs, where different prescriptions can even lead to qualitatively different conclusions. The proposed IPD framework provides a more self-consistent treatment of thermal corrections, offering improved theoretical control and more reliable predictions for phase transition dynamics and their GW signatures.


{\it \textbf{Construction of the Improved Partial Dressing.}} We seek a self-consistent resummation scheme that remains applicable over the full-temperature range. We therefore build on the PD approach, which directly resums daisy diagrams, and modify it to address the inconsistencies discussed below. The main features and limitations of the conventional prescriptions and the IPD scheme are summarized in Table~\ref{tab:scheme_summary} (see Sec.~\ref{sec:II} in the Supplemental Material~\cite{SM} for the detailed descriptions of the conventional prescriptions).

\begin{table*}[t]
\centering
\footnotesize
\caption{Comparison of commonly used thermal resummation schemes for the finite-temperature effective potential. For each scheme, the central idea, advantages, and limitations are summarized. The last row summarizes our new approach.}
\label{tab:scheme_summary}
\renewcommand{\arraystretch}{1.4} 
\begin{tabular}{|l|l|l|l|}
\hline
\parbox{0.1\linewidth}{\centering\vspace{4pt} Scheme \vspace{4pt}}
& \parbox{0.285\linewidth}{\raggedright\vspace{4pt} Main Features \vspace{4pt}}
& \parbox{0.285\linewidth}{\raggedright\vspace{4pt} Advantages \vspace{4pt}}
& \parbox{0.285\linewidth}{\raggedright\vspace{4pt} Limitations \vspace{4pt}} \\
\hline
\parbox{0.1\linewidth}{\centering\vspace{4pt} AE \vspace{4pt}}
& \parbox{0.285\linewidth}{\raggedright\vspace{4pt} Only resums zero-mode daisy diagrams \vspace{4pt}}
& \parbox{0.285\linewidth}{\raggedright\vspace{4pt} Computationally light; recovers high-temperature loop expansion \vspace{4pt}}
& \parbox{0.285\linewidth}{\raggedright\vspace{4pt} Relies on high-temperature approximation; strong scale dependence \vspace{4pt}} \\
\hline
\parbox{0.1\linewidth}{\centering\vspace{4pt} Parwani \vspace{4pt}}
& \parbox{0.285\linewidth}{\raggedright\vspace{4pt} Global thermal mass replacement across the one-loop potential \vspace{4pt}}
& \parbox{0.285\linewidth}{\raggedright\vspace{4pt} Simple to implement; recovers high-temperature loop expansion \vspace{4pt}}
& \parbox{0.285\linewidth}{\raggedright\vspace{4pt} Relies on high-temperature approximation; non-zero modes are not self-consistent \vspace{4pt}} \\
\hline
\parbox{0.1\linewidth}{\centering\vspace{4pt} DR \vspace{4pt}}
& \parbox{0.285\linewidth}{\raggedright\vspace{4pt} 3D effective theory via high-temperature matching \vspace{4pt}}
& \parbox{0.285\linewidth}{\raggedright\vspace{4pt} Formally systematic; less renormalization scale dependence \vspace{4pt}}
& \parbox{0.285\linewidth}{\raggedright\vspace{4pt} Requires strict scale separation; high-temperature dependence \vspace{4pt}} \\
\hline
\parbox{0.1\linewidth}{\centering\vspace{4pt} PD \vspace{4pt}}
& \parbox{0.285\linewidth}{\raggedright\vspace{4pt} Gap equation and derivative level substitution \vspace{4pt}}
& \parbox{0.285\linewidth}{\raggedright\vspace{4pt} Applicable at all temperatures; no double counting problem \vspace{4pt}}
& \parbox{0.285\linewidth}{\raggedright\vspace{4pt} Overlapping momentum problem; gap equation diverges at zero mass \vspace{4pt}} \\
\hline
\parbox{0.1\linewidth}{\centering\vspace{4pt} IPD (this work) \vspace{4pt}}
& \parbox{0.285\linewidth}{\raggedright\vspace{4pt} Features of the PD + Reclassified daisy diagrams and two-step resummation  \vspace{4pt}}
& \parbox{0.285\linewidth}{\raggedright\vspace{4pt} Advantages of the PD + No overlapping momentum problem; better consistency \vspace{4pt}}
& \parbox{0.285\linewidth}{\raggedright\vspace{4pt} Numerically more involved \vspace{4pt}} \\
\hline
\end{tabular}
\end{table*}

\begin{figure}[t]
    \centering
     \subfloat[]{
        \includegraphics[width=0.1\textwidth]{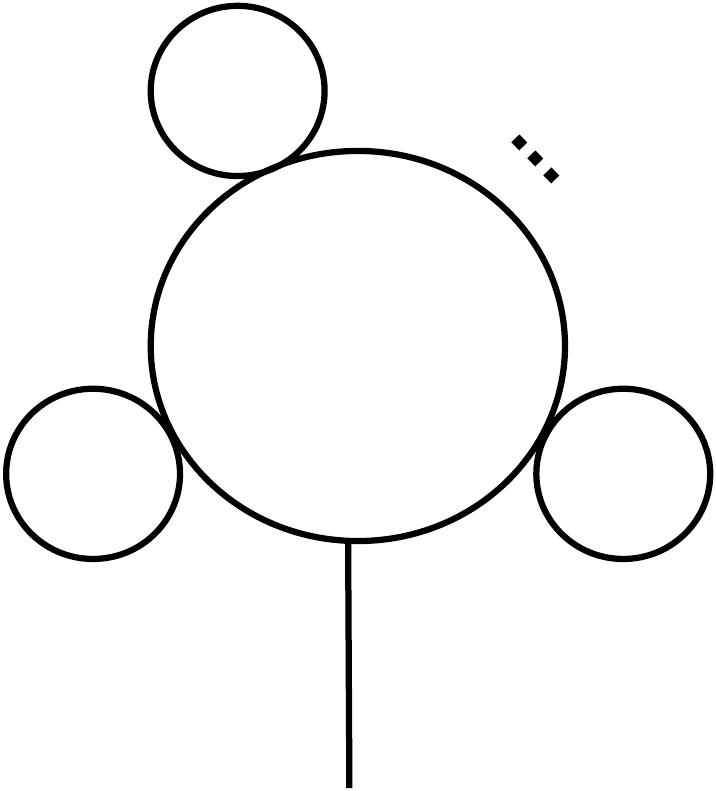}%
        \label{fig:PDdaisy_a}%
        }
    \quad
    \raisebox{0.5\height}{$\Longrightarrow$}
    \quad
    \subfloat[]{%
        \includegraphics[width=0.08\textwidth]{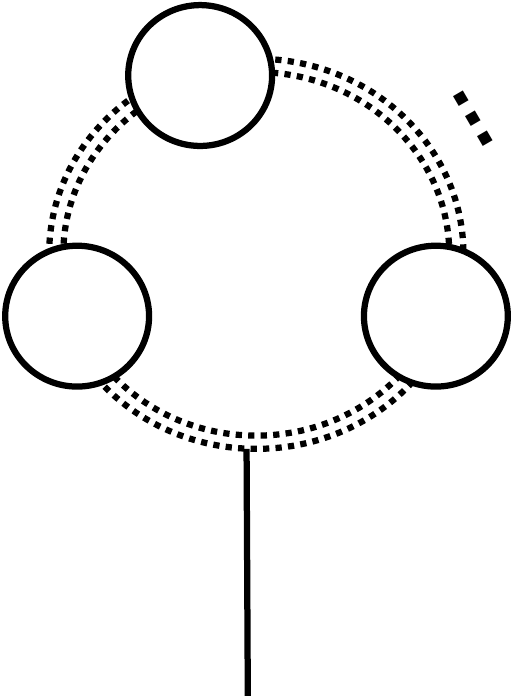}%
        \label{fig:PDdaisy_b}%
    }
    \caption{Classification and ordering of the tadpole resummation in the IPD scheme.}
    \label{fig:PDdaisy}
\end{figure}

We retain the external momentum dependence of the self-energy. The one-loop resummed tadpole diagram reads
\begin{align}
    V_{\rm tadpole}^i  = \frac{1}{2}g_i\frac{\partial m_i^2}{\partial \phi}\sint_K \frac{1}{K^2+m^2_i + \Pi_i(K)},
\end{align}
where we denote
\begin{align}
    \sint_K \equiv T\sum_{n=-\infty}^\infty\int\frac{\mathrm{d}^3k}{(2\pi)^3},\quad K^2 \equiv\omega_n^2 + k^2.\label{eq:sum-int}
\end{align}
Evaluating this sum-integral exactly is generally intractable, and appropriate approximations are therefore required. In the literature, the self-energy is often evaluated in the IR limit, \(\omega_n=0\) and \(\vec{k}\to0\). This approximation is self-consistent for the zero modes but not for the non-zero modes. Moreover, even for the zero modes, taking this limit can give rise to IR divergent contributions to the self-energy, such as \(\lambda\phi^2T/m\), which become singular as \(m^2\to0\). Since the non-zero modes are not responsible for the IR problem, their momentum-dependent self-energy contributions can therefore be neglected in the resummation. We first resum the daisy diagrams shown in Fig.~\ref{fig:PDdaisy_a}, and then consider the momentum-dependent self-energy contributions of the zero modes separately. 

As the first step, we consider the momentum-independent self-energy. The superdaisy contributions can then be incorporated into the resummation by solving the gap equation
\begin{equation}
\begin{aligned}
\adjustbox{valign=c, width=0.08\textwidth}{\includegraphics{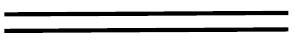}}
&=
 \adjustbox{valign=c, width=0.08\textwidth}{\includegraphics{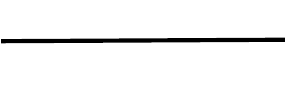}}
+
\adjustbox{valign=c, width=0.09\textwidth}{\includegraphics{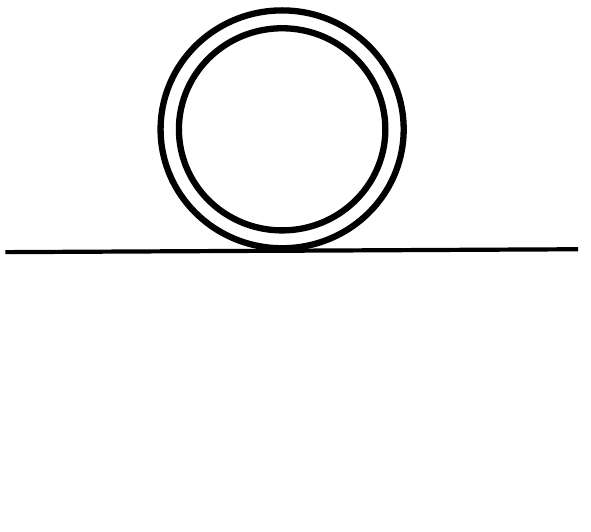}}
+
\adjustbox{valign=c, width=0.09\textwidth}{\includegraphics{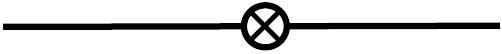}},
\\[-10pt]
M^2 &=m^2+\Pi_4(M^2)+\delta m^2,\label{mpdgeq}
\end{aligned}
\end{equation}
where the finite mass counterterm \(\delta m^2\) arises from renormalization and is explicitly included here.
This procedure is applicable to all Matsubara modes and does not introduce any IR singularities. The resulting effective mass \(M^2\) is then used in place of the tree-level mass \(m^2\) in the subsequent calculations. As the second step, the zero modes are resummed once more by incorporating the daisy diagrams shown in Fig.~\ref{fig:PDdaisy_b}. The resummed one-loop tadpole diagrams then take the form
\begin{align}
V^{i,R}_{\rm tadpole}=& \frac{1}{2}g_i\frac{\partial m^2_i}{\partial \phi}\left[
\sint_K^\prime \frac{1}{K^2+M^2_i}
\right. \nonumber\\
& \left. +T\int\frac{\mathrm{d}^3k}{(2\pi)^3}
\frac{1}{k^2+M^2_i+\Pi_{i,3}(k)}\right],\label{eq:1looptadpole}
\end{align}
where the prime on the sum-integral in the first term indicates that the \(n=0\) mode is omitted.
A complete evaluation of the momentum-dependent self-energy \(\Pi_3(k)\) is unnecessary, and suitable approximations can be made. For the non-zero modes, the IR limit \(\vec{k}\to0\) provides a good approximation, and the same approximation is valid for the contributions from heavy bosonic zero modes. Therefore, it is sufficient to retain the momentum dependence of the self-energy only for the light bosonic degrees of freedom that can give rise to IR problems. We thus have
\begin{align}
    \Pi_3(k) \simeq \Pi_3^{\rm heavy} (0) + \Pi_3^{\rm light}(k) + \mathcal{O}(k^2).
\end{align}
Here, light degrees of freedom are defined as those for which \(M^2(\phi,T)\simeq0\) over some range of \(\phi\) near the phase transition temperature. Finally, integrating over \(\phi\) gives the resummed effective potential,
\begin{align}
    V_{\rm eff}^{\rm IPD}(\phi, T) = V_0(\phi) + V_{\rm CT}(\phi) + \int_0^\phi\mathrm{d}\phi^\prime~\sum_iV^{i,R}_{\rm tadpole},
\end{align}
where the finite counterterm potential \(V_{\rm CT}\) is also explicitly included.

It is worth noting that when the mass satisfies \(M^2<0\), the effective potential and self-energy develop imaginary parts. These arise from the analytic continuation involved in evaluating the contour integrals and are commonly interpreted as indicating the instability of the corresponding field configuration~\cite{Weinberg:1987vp}. Following the prescription commonly adopted in the literature, we retain only the real parts of the loop contributions when evaluating the resummed effective potential and solving the gap equation. For the momentum-dependent self-energy, however, analytic continuation to negative squared mass renders the real part of \(\Pi_3^{\rm light}(k)\) discontinuous and the imaginary part diverges at \(k=2|M|\) (see Eqs.~\eqref{eq:mdepself_real} and \eqref{eq:mdepself_imag} in the Supplemental Material). Therefore, when calculating the integral of the second term on the right-hand side of Eq.~\eqref{eq:1looptadpole}, we keep the imaginary part of \(\Pi_3^{\rm light}(k)\).

{\it \textbf{Application to the Inert Doublet Model.}} The IDM provides a well-motivated benchmark for studying thermal phase transitions beyond the Standard Model (SM)~\cite{Deshpande:1977rw,Barbieri:2006dq,LopezHonorez:2006gr,Huang:2017rzf,Jiang:2022btc}. As a minimal extension of the SM with an additional scalar doublet protected by a discrete \(\mathbb{Z}_2\) symmetry, it provides a viable dark matter candidate~\cite{Barbieri:2006dq,LopezHonorez:2010eeh} and can accommodate a  FOPT in appropriate regions of parameter space~\cite{Chowdhury:2011ga}. We compare IPD with AE, Parwani, PD, and DR in IDM and assess the resulting differences in the transition parameters and GW spectra. The tree-level scalar potential is given by
\begin{align}
V(\Phi, \eta)=&\ \mu_{1}^{2}\Phi^\dag\Phi+\mu_{2}^{2}\eta^\dag\eta+\frac{1}{2}\lambda_{1}(\Phi^\dag\Phi)^{2}+\frac{1}{2}\lambda_{2}(\eta^\dag\eta)^{2} \nonumber\\
& +\lambda_{3}(\Phi^\dag\Phi)(\eta^\dag \eta)+\lambda_{4}(\Phi^{\dag}\eta)(\eta^\dag\Phi)\nonumber\\
&+\frac{1}{2}\{\lambda_{5}(\Phi^{\dagger}\eta)^{2}+\mathrm{h.c.}\},\label{VPhieta}
\end{align}
Here, \(\Phi\) is the SM Higgs doublet and \(\eta\) is the inert doublet. 
Boundedness of the scalar potential requires \(\lambda_1>0,\ \lambda_2>0,\
\lambda_3+\lambda_4-|\lambda_5|>-\sqrt{\lambda_1\lambda_2}\)~\cite{Barbieri:2006dq,LopezHonorez:2010eeh}. 
Expanding around the electroweak vacuum, the scalar doublets are parameterized as
\begin{align}
\Phi=\begin{pmatrix}G^+\\(\phi+h+iG^0)/\sqrt{2}\end{pmatrix},
\eta=\begin{pmatrix}H^+\\(H+iA)/\sqrt{2}\end{pmatrix},
\label{Phieta}
\end{align}
where \(\phi\) is the real scalar background field and \(h\) denotes the SM Higgs boson. The fields \(G^+\) and \(G^0\) are the charged and neutral Goldstone bosons, respectively. The inert scalar sector consists of the charged scalar \(H^+\), the CP-even scalar \(H\), and the CP-odd scalar \(A\).

At zero temperature, the background field takes the value of the electroweak vacuum expectation value  \(\phi=v_0=246.22~\mathrm{GeV}\). 
The additional inert scalar degrees of freedom can contribute to the electroweak oblique parameter \(T\). The corresponding contribution can be approximately expressed as
\begin{align}
    \Delta T \simeq \frac{1}{6\pi g^2 s_w^2 v_0^2}\big[m_{H^\pm} - m_H\big]\big[ m_{H^\pm} - m_A \big],
\end{align}
where \(s_w\) denotes the sine of the Weinberg angle. By assuming \(m_{H^\pm}(v_0)=m_A(v_0)\), corresponding to \(\lambda_4=\lambda_5\), one obtains \(\Delta T\simeq0\), indicating that the impact of the inert scalar sector on electroweak precision measurements becomes negligible. 
We choose the zero temperature tree-level masses of scalar particles \(\hat{m}_h\), \(\hat{m}_A\), \(\hat{m}_H\), \(\hat{m}_{H^\pm}\), and the couplings $\lambda_2$ and $\lambda_{L} = (\lambda_3+\lambda_4+\lambda_5)/2$ as the physical input parameters.
Furthermore, by choosing the renormalization scale as \(\mu_R=\hat{m}_Z\), where \(\hat{m}_Z\) denotes the physical pole mass of the \(Z\) boson, the SM physical parameters and their relations can be directly employed. 
The effective potentials for the IPD, AE, Parwani, and PD schemes are calculated in the Landau gauge using the modified minimal subtraction (\(\overline{\rm MS}\)) scheme. For the DR scheme, we use the package \texttt{DRalgo}~\cite{Ekstedt:2022bff,Bernardo:2026nyq,Fonseca:2020vke} to perform the DR and compute the effective potential. The detailed formulations and implementations of these resummation prescriptions are provided in the Supplemental Material.

With the resummed effective potential \(V_\text{eff}(\phi,T)\) obtained under different thermal resummation schemes, we proceed to extract the key dynamical parameters of the FOPT. The choice of thermal resummation modifies the shape of \(V_\text{eff}(\phi,T)\), including the depth of the vacua and the height of the potential barrier, which directly controls bubble nucleation and the subsequent cosmological evolution of the transition. 

\begin{table}[htbp]
    \centering
    \caption{Input and derived parameters for BP1.}
    \label{tab:BP1_input}
    \begin{tabular}{|c|c|c|c|}
    \hline
    \makecell{Input\\parameter} & \makecell{Value} & \makecell{Derived\\parameter} & \makecell{Value} \\
    \hline
    $\hat m_h$ & $125.10~\mathrm{GeV}$ & $\lambda_1$ & $0.26$ \\
    \hline
    $v_0$ & $246.22~\mathrm{GeV}$ & $\mu_1^2$ & $-7825~\mathrm{GeV}^2$ \\ \hline
    $\lambda_2$ & $0.2$ & $\mu_2^2$ & $4134~\mathrm{GeV}^2$ \\ \hline
    $\lambda_L$ & $0.0015$ & $\lambda_3$ & $5.14$ \\ \hline
    $\hat m_A$ & $400~\mathrm{GeV}$ & $\lambda_4$ & $-2.57$ \\ \hline
    $\hat m_{H^\pm}$ & $400~\mathrm{GeV}$ & $\lambda_5$ & $-2.57$ \\ \hline
    $\hat m_H$ & $65~\mathrm{GeV}$ & & \\
    \hline
    \end{tabular}

    \vspace{8pt} %

    \caption{Comparison of the phase transition parameters obtained using different resummation schemes for BP1, defined by Table~\ref{tab:BP1_input}. }
    \label{tab:scheme_compareBP1}
    \begin{tabular}{|c|c|c|c|c|c|c|}
    \hline
    Scheme
    & $T_c\,(\mathrm{GeV})$
    & $T_n\,(\mathrm{GeV})$
    & $\xrightarrow[\rm type]{\rm Transition}$
    & $v_n/T_n$
    & $\alpha$
    & $\beta/H$ \\
    \hline
    Parwani & 88.20 & 78.86 & FOPT & 2.75 & 0.0341 & 1888 \\
    \hline
    PD & 93.23 & 77.71 & FOPT & 3.27 & 0.0566 & 1064 \\
    \hline
    AE & 112.28 & 100.17 & FOPT & 2.28 & 0.0148 & 1811 \\
    \hline
    DR & \multicolumn{6}{c|}{Symmetry non-restoration at high temperature} \\
    \hline
    IPD & 96.01 & 79.97 & FOPT & 2.96 & 0.0503 & 1079 \\
    \hline
    \end{tabular}
\end{table}

The key phase transition parameters directly extracted from the effective potential are the critical temperature \(T_c\), the nucleation temperature \(T_n\), the phase transition strength \(\alpha\), the ratio of the true vacuum field value to temperature at nucleation \(v_n/T_n\) and the ratio of characteristic timescale to the Hubble parameter \(\beta/H\). 
The parameters \(\alpha\) and \(\beta/H\) characterize the available energy released during the transition and the duration of the phase transition, respectively, and serve as the primary inputs for the calculations of GW energy density spectrum \(\Omega_{\rm GW}h^2\). We use the package \texttt{PhaseTracer}~\cite{Athron:2020sbe,Athron:2024xrh} to evaluate these phase transition parameters.

\begin{figure}[t]
    \centering    \includegraphics[width=0.32\linewidth]{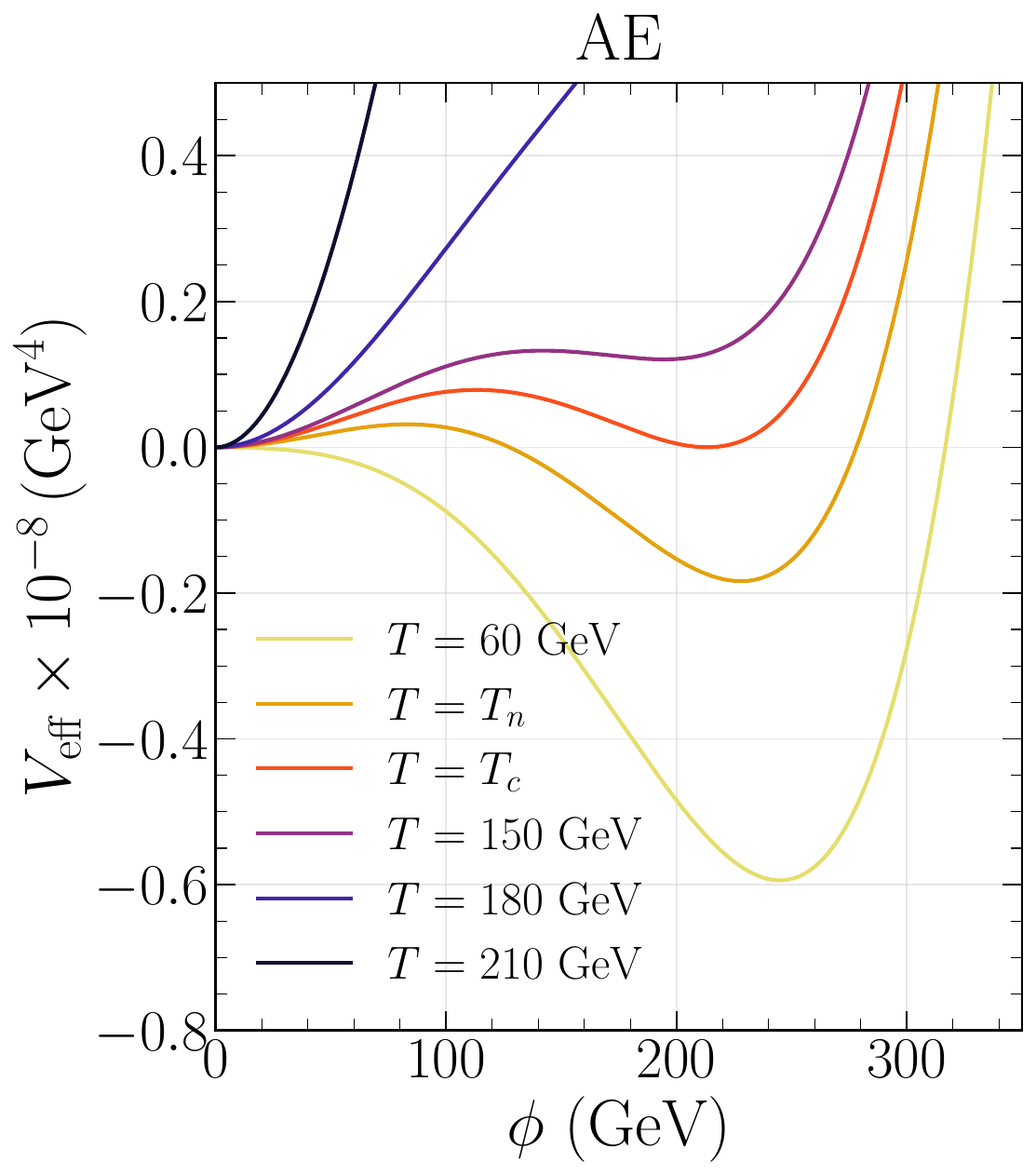}
    \includegraphics[width=0.32\linewidth]{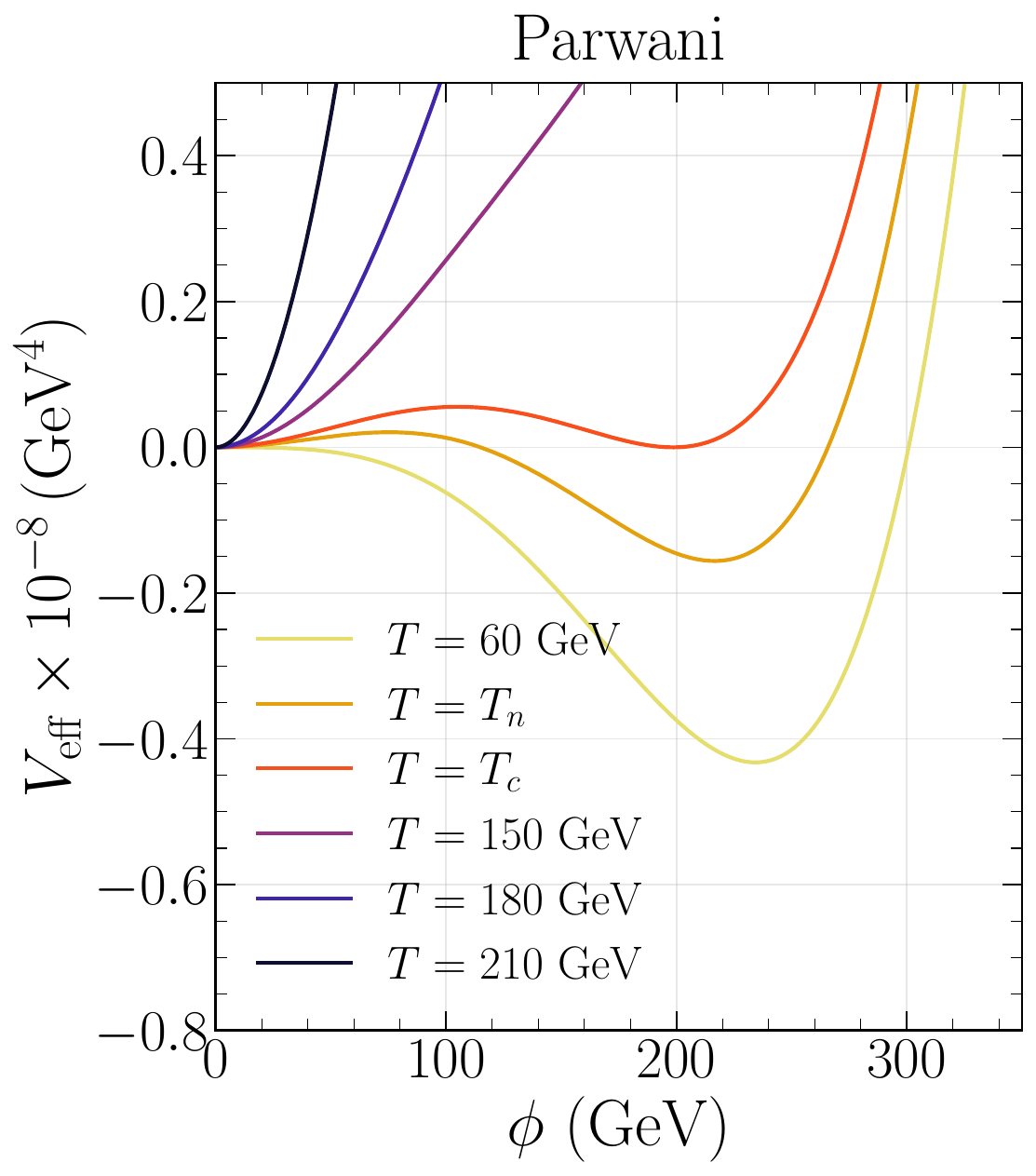}
    \includegraphics[width=0.32\linewidth]{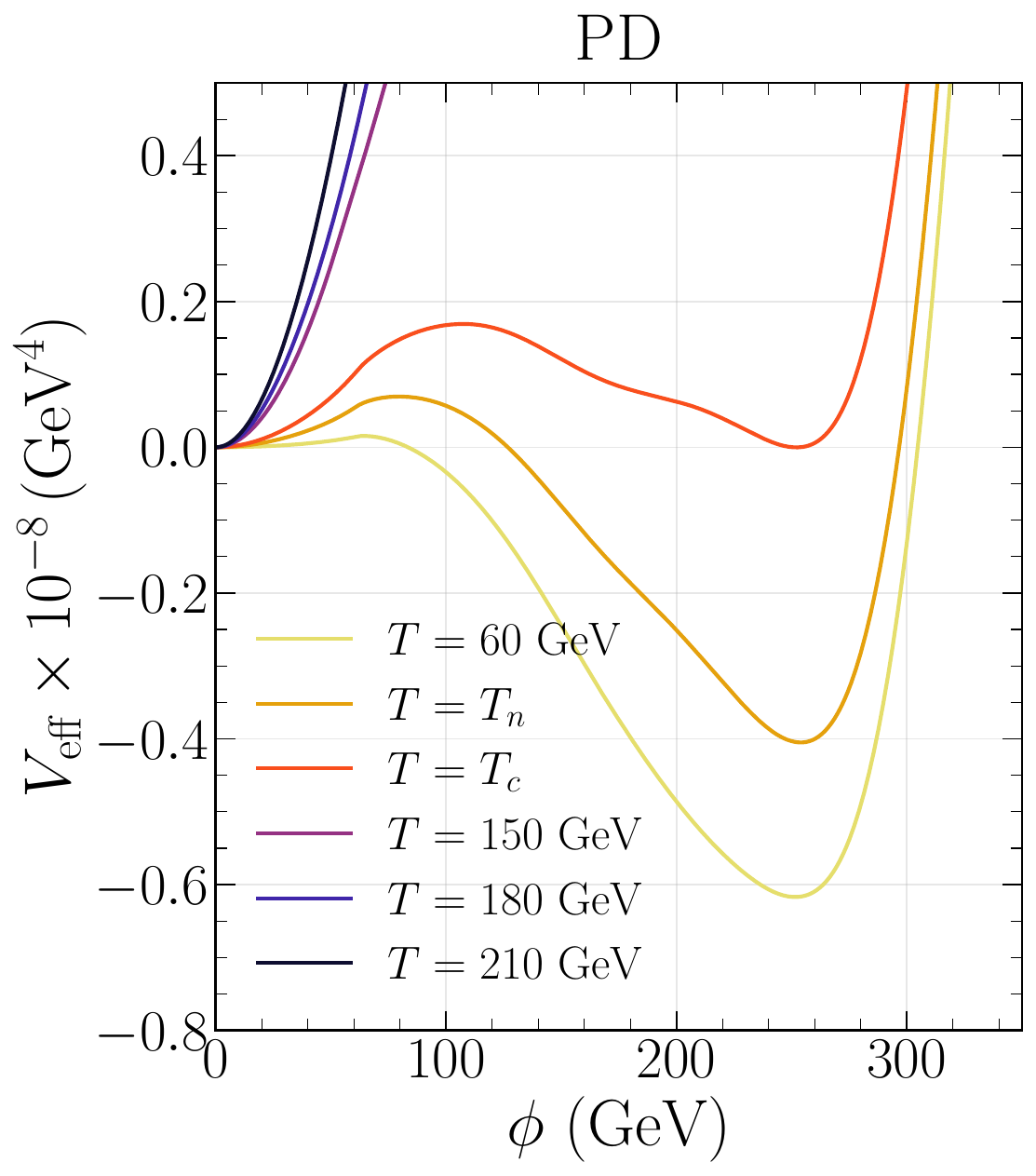} 
    \includegraphics[width=0.32\linewidth]{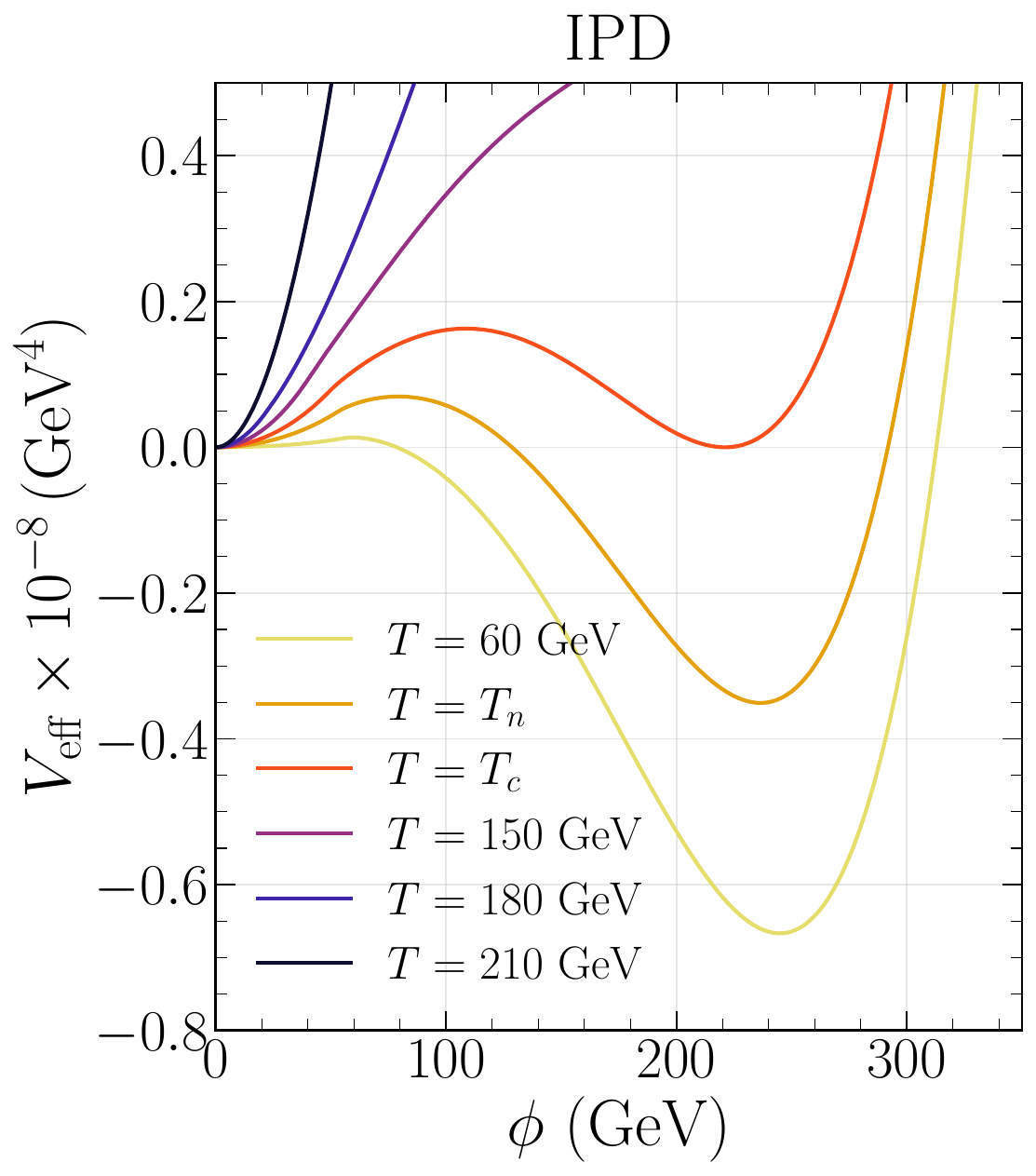}
    \includegraphics[width=0.32\linewidth]{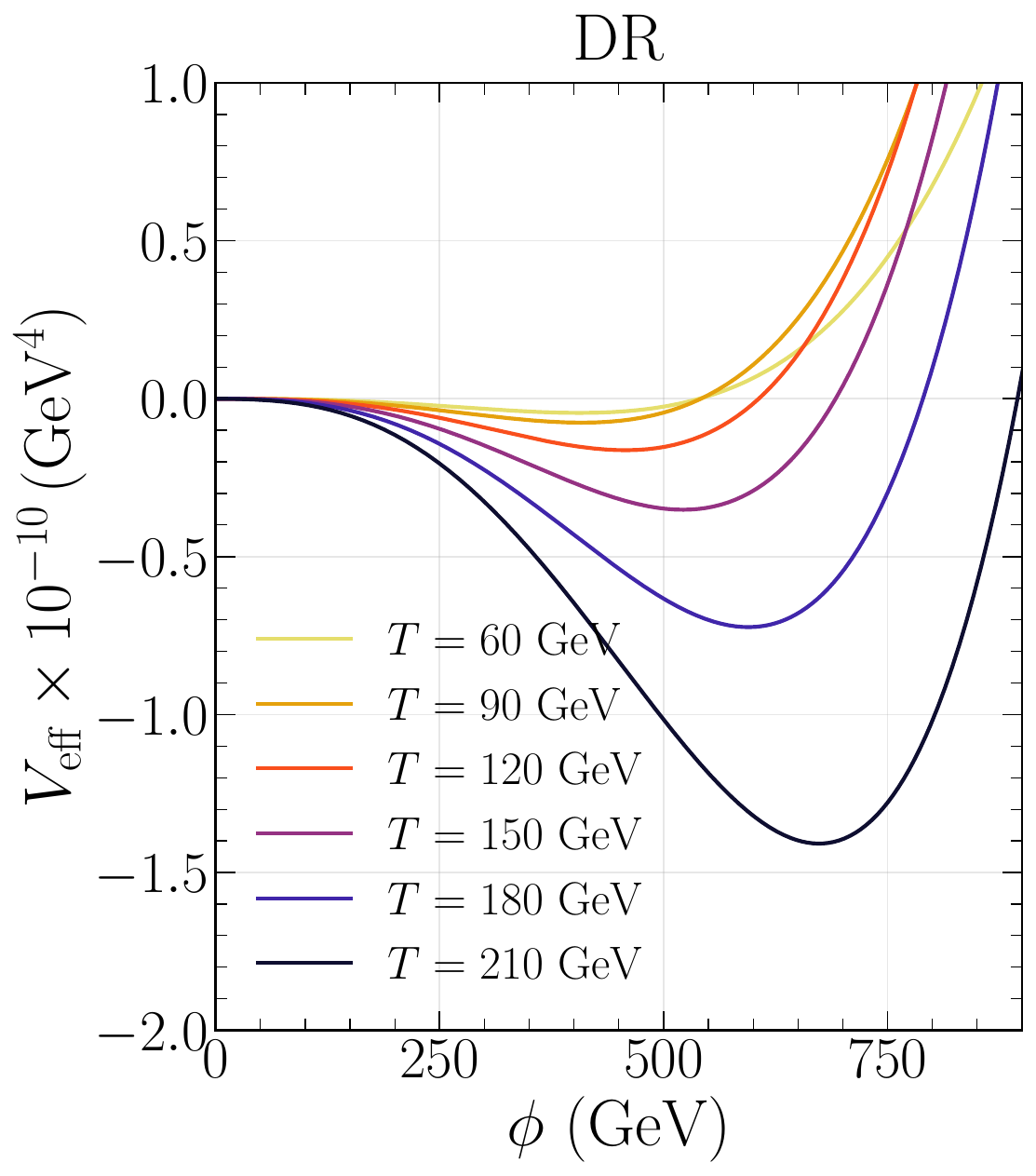}
    \caption{Comparison of the effective potential for  BP1, defined in Table~\ref{tab:BP1_input}, across five resummation schemes: AE (top left), Parwani (top middle), PD (top right), IPD (bottom left), and DR (bottom right). Different prescriptions yield distinct potential behaviors, including variations in barrier height and critical temperature, as well as symmetry non-restoration for the DR scheme. The corresponding phase transition parameters are collected in Table~\ref{tab:scheme_compareBP1}.}
    \label{fig:comparePotential}
\end{figure}

\begin{figure}
    \centering
    \includegraphics[width=0.7\linewidth]{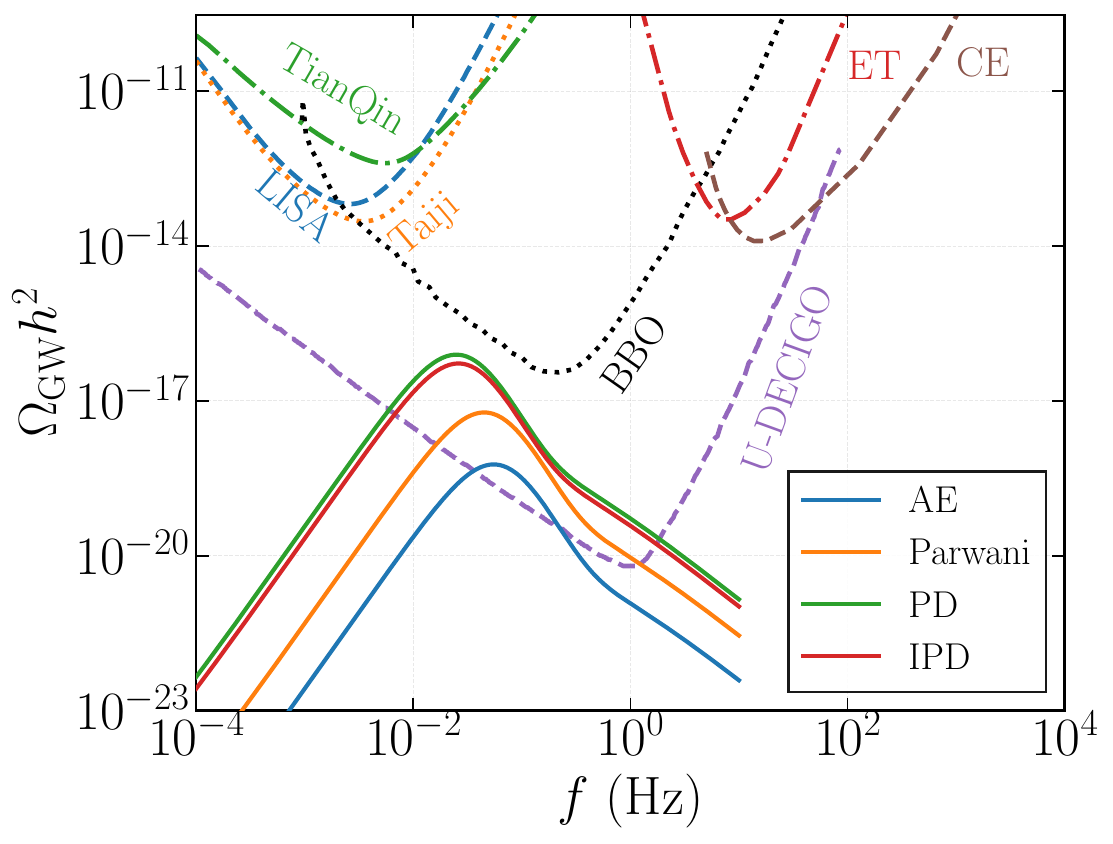}
    \caption{GW energy spectra for BP1, illustrated against the experimental sensitivity curves of various proposed GW detectors such as LISA, TianQin, Taiji, BBO, CE, ET and U-DECIGO. The blue, orange, green, and purple solid lines indicate the GW energy density considering the AE, Parwani, PD, and IPD prescriptions, respectively.  }
    \label{fig:GW}
\end{figure}

We consider a benchmark point in the IDM, denoted by BP1, with the input parameters specified in Table~\ref{tab:BP1_input}. The phase transition parameters obtained using different resummation schemes are summarized in Table~\ref{tab:scheme_compareBP1}, while Fig.~\ref{fig:comparePotential} shows the temperature evolution of the corresponding effective potentials. 
Notably, the DR scheme exhibits non-restoration of symmetry at high temperatures, unlike the other schemes considered here. This unphysical symmetry non-restoration behavior generally signals the failure of the DR scheme within this parameter region. In this case, more sophisticated treatments of the three-dimensional effective field theory may be required~\cite{Kierkla:2025qyz}.

Furthermore, the AE and Parwani schemes, which rely on the high-temperature expansion, yield a lower potential barrier compared with the more self-consistent PD and IPD schemes, resulting in a weaker phase transition strength, as shown in Fig.~\ref{fig:comparePotential} and Table~\ref{tab:scheme_compareBP1}. For the PD scheme, the gap equation breaks down near Higgs effective mass \(M_h^2\simeq 0\), which renders the potential insufficiently smooth around the barrier and ultimately affects the numerical computation of phase transition dynamics.

Our proposed IPD resummation scheme improves upon the PD prescription, resulting in a more self-consistent and reliable framework. Although it involves a more numerically involved treatment, it yields more accurate phase transition parameters, which in turn lead to more precise predictions for the GW signature. Figure~\ref{fig:GW} shows the sum of three contributions from bubble collisions, sound waves, and magnetohydrodynamic turbulence for BP1 in different resummation prescriptions. The peak frequency and amplitude for the AE prescription are \(f_{\rm peak} = 0.054\) Hz and \(\Omega_{\rm GW} h^2_{\rm peak} = 5.9 \times 10^{-19} \); for Parwani, \(f_{\rm peak} = 0.045\) Hz and \(\Omega_{\rm GW} h^2_{\rm peak} = 6.0 \times 10^{-18}\); for PD, \(f_{\rm peak} = 0.025\) Hz and \(\Omega_{\rm GW} h^2_{\rm peak} = 7.9 \times 10^{-17}\); and for IPD, \(f_{\rm peak} = 0.026\) Hz and \(\Omega_{\rm GW} h^2_{\rm peak} = 5.3 \times 10^{-17}\).

{\it \textbf{Conclusions.}} We have developed the IPD prescription to address two limitations of the conventional PD organization: the treatment of external-momentum dependence in self-energy insertions and the singular behavior of the gap equation near vanishing effective mass.
By incorporating momentum-dependent contributions for IR-sensitive bosonic zero modes within a self-consistent resummation framework, IPD provides an more self-consistent framework for thermal resummation beyond the high-temperature approximation. We have implemented IPD in the IDM and compared its predictions with the AE, Parwani, DR, and conventional PD prescriptions. Our comparison in the IDM demonstrates that the choice of resummation prescription can significantly affect the predicted phase transition parameters and GW spectra. These results highlight the importance of a theoretically controlled treatment of thermal fluctuations for reliable predictions of cosmological phase transitions.

Within the viable IDM parameter space considered here, the transitions are generally weak, and the differences between the IPD and PD GW spectra are consequently less pronounced than the differences in the transition parameters. It would therefore be interesting to extend the comparison to models exhibiting stronger FOPTs, where the effects of different resummation prescriptions on the predicted GW signals may be more pronounced. More broadly, a theoretically consistent treatment of thermal fluctuations is essential for obtaining reliable predictions for GW signals and related phase transition phenomena, including electroweak baryogenesis, dark matter production, primordial black hole formation, and primordial magnetic field generation. It is also important for reliably reconstructing particle physics model parameters from future GW observations.

{\it \textbf{Acknowledgments.}} We thank Siyu Jiang for the helpful discussions. This work is supported by the National Natural Science Foundation of China (NNSFC) under Grant No.12475111.


\bibliographystyle{apsrev4-1}
\bibliography{ref}


\include{Supp_for_PRL}


\end{document}

%% file: Supp_for_PRL.tex

\clearpage
\newpage
\maketitle
\onecolumngrid
\begin{center}
\textbf{\large Towards Consistent Thermal Resummation: Improved Partial Dressing for Precise Electroweak Phase Transition Predictions} \\ 
\vspace{0.05in}
{\it \large Supplementary Material}\\ 
\vspace{0.05in}
{Dayun Qiu and Fa Peng Huang}
\end{center}
\onecolumngrid

\setcounter{equation}{0}
\setcounter{figure}{0}
\setcounter{table}{0}
\setcounter{section}{0}
\makeatletter
\renewcommand{\theequation}{S\arabic{equation}}
\renewcommand{\thefigure}{S\arabic{figure}}
\renewcommand{\thetable}{S\arabic{table}}
\renewcommand{\thesection}{S\arabic{section}}
\renewcommand{\thesubsection}{\Alph{subsection}}
\renewcommand{\thesubsubsection}{\Roman{subsubsection}}
\makeatother

\tableofcontents

\section{Thermal Resummation}\label{sec:II}
In this section, we review commonly used thermal resummation prescriptions and formulate them within a common framework. This allows us to compare their perturbative structures and identify theoretical issues arising from conventional treatments. Our discussion clarifies which aspects of conventional resummation procedures require further consideration, thereby motivating the construction of IPD.

\subsection{Finite-temperature effective potential}\label{sec:ftep}
For a theory with scalar, fermionic, and vector fields, we expand the scalar fields about a classical background, \(\hat{\phi}=\phi+\delta\hat{\phi}\), and integrate out the quantum fluctuations. The resulting effective potential is organized as a loop expansion of one-particle-irreducible (1PI) vacuum diagrams. We work in Landau gauge.

At zero temperature, the one-loop effective potential is given by~\cite{Coleman:1973jx, Jackiw:1974cv}
\begin{align}
    V_{\rm CW}(\phi) = -\frac{1}{2}\ 
  \adjustbox{valign=c, width=0.1\textwidth}{   \includegraphics{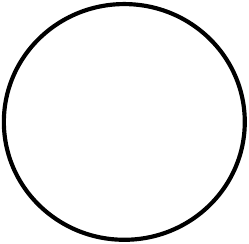}  } = -\frac{i}{2}\int\frac{\mathrm{d}^4k}{(2\pi)^4}\log\det\left( iG^{-1}\right),\label{1-loop}
\end{align}
where \(G^{-1}\) and ``\(\det\)'' represent the inverse propagator and functional determinant, respectively. In the \(\overline{\rm MS}\) scheme, the Coleman–Weinberg potential reads
\begin{align}
V_{\rm CW}(\phi)
= \sum_i (-1)^{s_i} g_i
\frac{m_i^4(\phi)}{64\pi^2}
\left[
\log\left(\frac{m_i^2(\phi)}{\mu_R^2}\right) - c_i
\right].\label{eq:CWpotential}
\end{align}
Here, \(s_i=1\) (\(0\)) for fermions (bosons), while \(c_i=3/2\) for scalars, fermions, and longitudinal gauge boson modes, and \(c_i=1/2\) for transverse gauge boson modes. The quantities \(g_i\) and \(m^2_i(\phi)\) denote the corresponding number of degrees of freedom and the field-dependent masses. \(\mu_R\) represents the renormalization scale.

At finite temperature \(T\), the effective potential receives an additional thermal contribution, which can be evaluated in the imaginary-time formalism~\cite{Matsubara:1955ws,Quiros:1999jp,Mustafa:2022got}. The temporal momentum component is then discretized as
\begin{align}
    k^0=i\omega_n,
\qquad
\omega_n=
\begin{cases}
2n\pi T, & \text{for bosons},\\
(2n+1)\pi T, & \text{for fermions},
\end{cases}
\end{align}
leading to the replacement
\begin{align}
\int\frac{\mathrm{d}^4 k}{(2\pi)^4}
\quad\longrightarrow\quad
iT\sum_{n=-\infty}^{\infty}
\int\frac{\mathrm{d}^3 k}{(2\pi)^3}.
\end{align}
Thus, the effective potential at finite temperature takes the form
\begin{align}
V_{\rm eff}(\phi,T)
= V_0(\phi) + V_{\rm CW}(\phi) + V_T(\phi,T),
\end{align}
where the third term \(V_T\) is the one-loop thermal correction,
\begin{align}
    V_T(\phi,T) = \sum_i(-1)^{s_i}g_i\frac{T^4}{2\pi^2}J_{b/f}\left(\frac{m^2_i(\phi)}{T^2}\right).
\end{align}
The thermal functions are defined as
\begin{align}
J_{b/f}(x^2)
= \int_{0}^{\infty} \mathrm{d}y~ y^2
\log\left(1\mp e^{-\sqrt{x^2+y^2}}\right).
\end{align}
In the high-temperature limit, corresponding to \(x^2\to 0\), these functions admit the expansions
\begin{align}
    J_b(x^2) \approx&\  -\frac{\pi^4}{45}
+ \frac{\pi^2}{12}x^2
- \frac{\pi}{6}x^{3}
- \frac{x^4}{32}\log\frac{x^2}{a_b}
+ \mathcal{O}(x^6),\\
J_f(x^2) \approx&\  \frac{7\pi^4}{360}
- \frac{\pi^2}{24}x^2
- \frac{x^4}{32}\log\frac{x^2}{a_f}
+ \mathcal{O}(x^6),
\end{align}
where \(a_b = 16\pi^2 \exp\left(3/2-2\gamma_E\right)\) and \( a_f = \pi^2 \exp\left(3/2-2\gamma_E\right)\). The \(x^3\) term originates from the bosonic Matsubara zero mode (\(n=0\)), which is responsible for the IR sensitivity of thermal field theory. 

\begin{figure}
    \centering
    \includegraphics[width=0.4\linewidth]{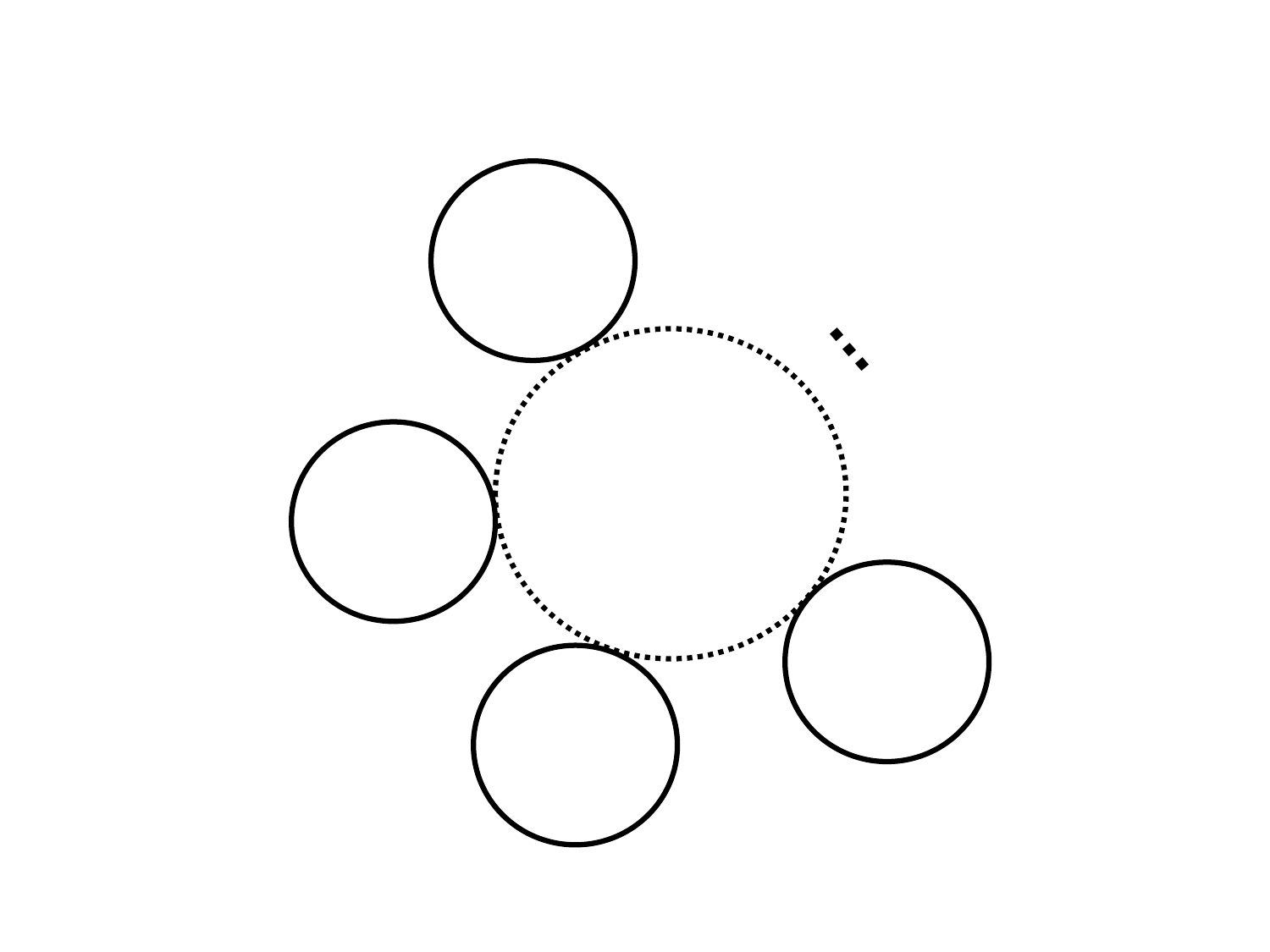}
    \caption{The bosonic zero-mode daisy diagrams dominating the leading IR divergence.}
    \label{fig:daisy}
\end{figure}

At higher orders, bosonic zero-mode daisy diagrams are increasingly IR sensitive (Fig.~\ref{fig:daisy}). In the high-temperature limit, a contribution with \(l\) self-energy insertions scales as \(T(m^2)^{3/2-l}\), and is IR divergent for \(l\geq2\) as \(m^2\to0\). Correspondingly, the ratio of successive loop orders scales as \(\lambda T^2/m^2\). When this ratio is not small, the ordinary loop expansion is not reliable and the daisy contributions must be resummed.

\subsection{Daisy resummation}
The IR problem can be addressed by resumming the thermally enhanced higher-loop contributions associated with the bosonic zero modes. In particular, the daisy (or ring) diagrams form an infinite class of IR-sensitive contributions that can be systematically incorporated through thermal mass corrections. Different resummation prescriptions differ in how these thermal corrections are implemented in the effective potential.

\subsubsection{Parwani scheme}
The Parwani resummation prescription is a scheme for recovering the perturbative expansion~\cite{Parwani:1991gq}. This scheme incorporates the thermal masses into the original Lagrangian,
\begin{align}
\mathcal{L} = \left(\mathcal{L} - \frac{1}{2} m_{\rm th}^2 \hat{\phi}^2 - \frac{1}{2} m^2_{D} A^2_0\right) +\frac{1}{2} m_{\rm th}^2 \hat{\phi}^2 + \frac{1}{2} m^2_{D} A^2_0 ,
\end{align}
while the same thermal mass terms are subtracted to ensure that the resulting Lagrangian remains equivalent to the original theory. The thermal mass terms in the first bracket are treated non-perturbatively, while the remaining terms are treated perturbatively (also known as thermal counterterms). For vector bosons, only the longitudinal modes (\(A_0\)) acquire thermal masses. Since fermions do not suffer from the IR problem, their thermal masses are not included in the resummation.

Then, using the background-field method to calculate the effective potential, the field-dependent masses are replaced by the corresponding effective masses, \(m_i^2(\phi)\rightarrow m_{i,\mathrm{eff}}^2(\phi,T)
= m_i^2(\phi)+ m_{\mathrm{th},i}^2.\) The thermal mass \(m_{\mathrm{th},i}^2\) is generally obtained from the limit of the self-energy, \(\Pi_i(K)\), namely \(\Pi_i(K\rightarrow0)\), where \(K\) denotes the Euclidean four-momentum. In the high-temperature limit,
\begin{align}
m_{\mathrm{th},i}^2=\Pi_i(0)\sim\lambda_i T^2.
\end{align}
After including the corresponding thermal counterterms, the effective perturbative expansion parameter scales as 
\begin{align}
   \frac{\lambda T}{\pi m_{\mathrm{eff}}} \sim \frac{\sqrt{\lambda}}{\pi} .
\end{align}
Thus, the perturbative expansion of the effective potential becomes well-defined again in the high-temperature regime. At one-loop order, the Parwani scheme implements the resummation by replacing the field-dependent masses with the corresponding effective masses, throughout the effective potential, including both the zero temperature Coleman--Weinberg and finite-temperature contributions:
\begin{align}
    V^{\rm Par}_{\rm eff}(\phi,T)
= V_0(\phi)
+ V_{\rm CW}(m^2_{\rm eff}(\phi,T))
+ V_T(m^2_{\rm eff}(\phi,T),T).
\end{align}

However, it should be noted that the \(K\rightarrow0\) limit used in the self-energy calculation is not applicable to non-zero Matsubara modes, for which \(K^2\sim T^2\). The approximation \(\Pi(0)\approx\Pi(K)\) is only justified when \(K\ll T\). Furthermore, for a scalar potential that induces spontaneous symmetry breaking, such as
\begin{align}
V_0 = \frac{1}{2}\mu^2\phi^2 + \frac{1}{8}\lambda\phi^4,\label{eq:V0}
\end{align}
the tree-level mass parameter satisfies \(\mu^2<0\). A valid perturbative expansion, therefore, requires \(\lambda T^2\gg-\mu^2\). But near the phase transition temperature, one typically has \(T^2\sim-\mu^2\), causing the perturbative expansion to break down again. This issue is absent in the massless pure scalar theory considered by Parwani. In this case, the scalar self-energy has no dependence on the external momentum, and the perturbative expansion does not break down again. However, when the Parwani prescription is extended to more general theories, the above problems can arise.

\subsubsection{Arnold--Espinosa scheme}
Another approach to restoring the perturbative expansion is the AE scheme~\cite{Arnold:1992rz}. In this scheme, only the bosonic zero-mode daisy diagrams, as shown in Fig.~\ref{fig:daisy}, are resummed, while the daisies involving non-zero modes can be treated perturbatively, order by order, as their perturbative expansion remains valid. The Lagrangian can be written as
\begin{align}
    \mathcal{L} = \left(\mathcal{L} - \frac{1}{2} m^2_{\rm th}\hat{\phi}^2_{n=0} - \frac{1}{2} m^2_D A^2_{0,n=0} \right) +  \frac{1}{2} m^2_{\rm th}\hat{\phi}^2_{n=0} + \frac{1}{2} m^2_D A^2_{0,n=0} .
\end{align}
As in the Parwani scheme, the last two terms are treated as thermal counterterms, but in the AE scheme, they are introduced only for the bosonic zero modes. 

Since the IR-sensitive contributions originate solely from the bosonic zero modes, the perturbative expansion parameter at high temperature is likewise dominated by the zero-mode daisy diagrams. Once the thermal masses are incorporated into the zero modes, the perturbative loop expansion is restored, as in the Parwani scheme.
At one-loop order, the AE resummation scheme is implemented by replacing \(m^2_i(\phi)\) with \(m^2_{i,\rm eff}(\phi,T)\) only in the cubic mass term of the effective potential,
\begin{align}
    V^{\rm AE}_{\rm eff}(\phi,T) = V_0(\phi) + V_{\rm CW}(m^2(\phi)) + V_T (m^2(\phi),T) + V_{\rm daisy}(\phi,T),
\end{align}
where
\begin{align}
    V_{\rm daisy}(\phi,T) = \sum_i-\frac{g_iT}{12\pi}\left(m^3_{i,\rm eff}(\phi,T) - m^3_i(\phi) \right).
\end{align}

Compared with the Parwani scheme, the AE scheme does not involve the self-energies of non-zero modes, and hence the limit \(\Pi(k^0=0;\vec{k}\to0)\) can be taken consistently.
For the theory defined in Eq.~\eqref{eq:V0}, however, this scheme remains applicable only in the high-temperature regime, \(\lambda T^2\gg-\mu^2\), and the perturbative expansion again breaks down near the phase transition temperature. In addition, the AE scheme exhibits substantially stronger renormalization-scale sensitivity than the Parwani prescription~\cite{Curtin:2016urg,Curtin:2022ovx}. This difference arises because Parwani’s global mass replacement induces partial cancellation of scale-dependent terms across the zero-temperature and finite-temperature sectors, an effect absent from the zero-mode-only AE prescription.

\subsubsection{Dimensional reduction}
At high temperatures, thermal field theory exhibits a hierarchy of scales, 
\begin{align}
    \pi T \gg gT \gg g^2T,
\end{align}
where \(g\) is the coupling
constant at the scale \(\pi T\). They correspond to the hard, soft, and ultrasoft scales, respectively.
The DR exploits this hierarchy by successively integrating out the higher-energy scales and constructing effective theories for the remaining lower-energy degrees of freedom~\cite{Farakos:1994kx,Kajantie:1995dw,Braaten:1995cm}, which provides a formally more systematic perturbative framework and reduces the renormalization-scale dependence of the resulting predictions~\cite{Croon:2020cgk,Schicho:2021gca}. 
The first step is to integrate out the non-zero Matsubara modes at the hard scale \(\pi T\), yielding a three-dimensional effective theory for the bosonic zero modes. Depending on the desired accuracy and the scale of interest, the soft scale \(gT\) can be integrated out further, leaving an effective theory describing the ultrasoft dynamics at the scale \(g^2T\). For a scalar field, the resulting effective Lagrangian can be written as
\begin{align}
\mathcal{L}_{3d}
=
\frac{1}{2}(\nabla\phi_{3d})^2
+\frac{1}{2}m_{3d}^2\phi_{3d}^2
+\frac{\lambda_{3d}}{4!}\phi_{3d}^4+\cdots ,
\end{align}
where \(m_{3d}^2\) and \(\lambda_{3d}\) are obtained by matching the three-dimensional theory to the original four-dimensional theory. At leading order,
\begin{align}
m_{3d}^2 = m_{4d}^2 + \Pi_{\rm hard}(0),\qquad
\lambda_{3d} = \lambda T+\cdots ,
\end{align}
where \(\Pi_{\rm hard}\) denotes the contribution from the non-zero Matsubara modes. The four-dimensional and three-dimensional fields are related by
\begin{align}
\phi_{3d}^2 = \frac{1}{T}\left(1+\Pi_{\phi}^\prime(0)-\delta Z_\phi\right)\phi_{4d}^2,
\end{align}
where \(\Pi_\phi^\prime(k)\) denotes the derivative of \(\Pi_\phi(k)\) with respect to \(k^2\), and \(\delta Z_\phi\) denotes the corresponding field-renormalization counterterm. The IR sensitive dynamics are then contained entirely in the three-dimensional theory, with the bosonic zero modes treated explicitly, while the effects of the non-zero modes are incorporated through perturbative matching. Up to one-loop order, the effective potential reads
\begin{align}
    V_{3d,\rm eff}(\phi_{3d}, T) = V_{3d,0} + V_{3d,1}, 
\end{align}
where the tree-level potential $V_{3d,0}$ takes the same form as its four-dimensional counterpart, with all parameters replaced by their three-dimensional versions. The one-loop correction is given by
\begin{align}
    V_{3d,1}(\phi_{3d},T) = - \frac{1}{12\pi}\sum_i g_i m^3_{i}(\phi_{3d}).
\end{align}

The DR framework relies on a hierarchy of thermal scales and, consequently, on a weak-coupling and high-temperature expansion. Its accuracy can therefore deteriorate when the separation between the scales becomes less pronounced or when the relevant masses are no longer small compared with \(T\). In addition, after integrating out the hard and soft scales, the remaining ultrasoft scale \(g^2T\) can become intrinsically non-perturbative, particularly in non-Abelian gauge theories, requiring non-perturbative treatment of the resulting three-dimensional effective theory~\cite{Braaten:1995jr}.

\subsubsection{Partial dressing}
\begin{figure}[t]
    \centering    \includegraphics[width=0.3\linewidth]{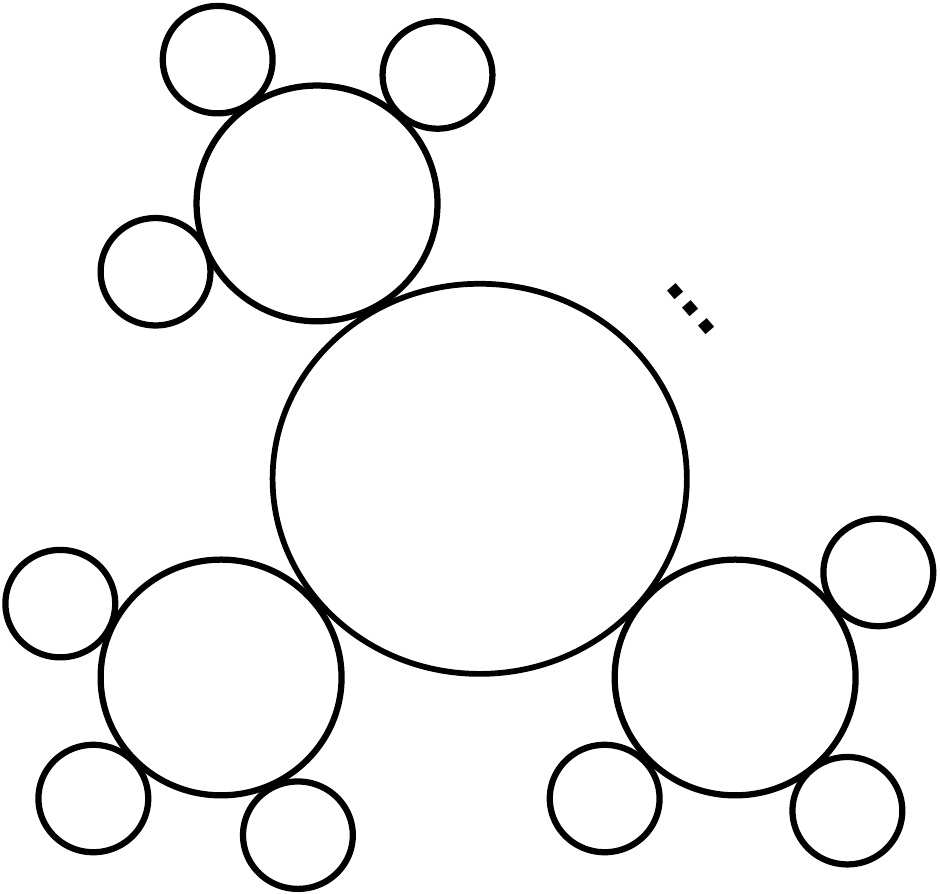}
    \caption{The superdaisy contributions.}
    \label{fig:superdaisy}
\end{figure}
\begin{figure}[t]
    \centering
    \subfloat[]{%
        \includegraphics[width=0.2\textwidth]{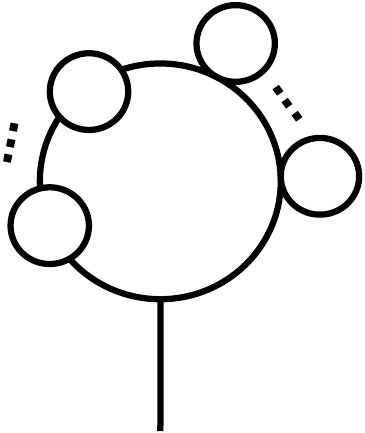}%
        \label{fig:FD_a}%
    }
    \hfill
    \subfloat[]{%
        \includegraphics[width=0.2\textwidth]{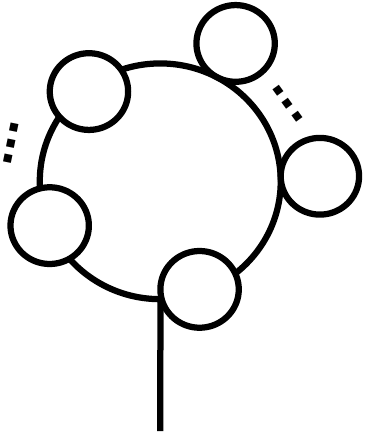}%
        \label{fig:FD_b}%
    }
    \hfill
    \subfloat[]{%
        \includegraphics[width=0.2\textwidth]{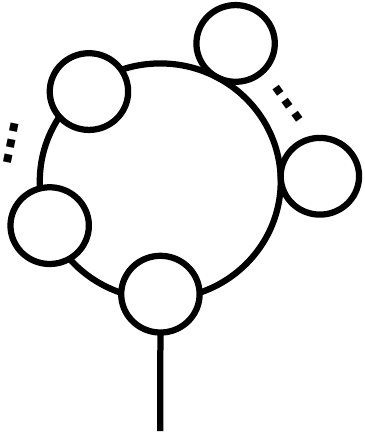}%
        \label{fig:FD_c}%
    }
    \hfill
    \subfloat[]{%
        \includegraphics[width=0.2\textwidth]{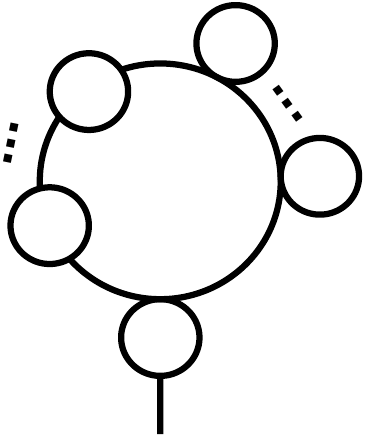}%
        \label{fig:FD_d}%
    }
    \caption{The tadpole diagrams.}
    \label{fig:FD}
\end{figure}

Another class of resummation methods directly resums the daisy diagrams in the theory shifted around the background field. In general, the superdaisy diagrams, as shown in Fig.~\ref{fig:superdaisy}, are also included by solving the gap equation
\begin{align}
M^2 = m^2 + \Pi(M^2,T).\label{eq
:gap}
\end{align}
Here, the external momentum of the self-energy \(\Pi\) is likewise set to zero. A representative approach is the tadpole method, in which the one-loop tadpole diagrams, as shown in Fig.~\ref{fig:FD}, are resummed to obtain the first derivative of the effective potential, \(\frac{\mathrm{d}V_{\rm eff}}{\mathrm{d}\phi}\). 

Figure~\ref{fig:FD_a} only involves propagator dressing, whereas Figs.~\ref{fig:FD_b}, \ref{fig:FD_c}, and \ref{fig:FD_d} also involve three-point vertex dressing. Resumming only the contribution shown in Fig.~\ref{fig:FD_a} is referred to as ``partial dressing''~\cite{Boyd:1992xn}, while including the additional vertex dressing is referred to as ``full dressing''~\cite{Espinosa:1992gq,Espinosa:1992kf}. Including the superdaisy diagrams, the full dressing scheme amounts to inserting the effective mass \(M^2\), obtained by solving the gap equation, into the one-loop effective potential. However, as pointed out in Ref.~\cite{Boyd:1993tz}, the full dressing scheme overcounts the two-loop daisy diagrams. This can be seen directly from Fig.~\ref{fig:FD}: at two-loop order, the same daisy diagram is generated by the four tadpole diagrams, and resumming all of them leads to overcounting of the same two-loop contribution. The PD scheme avoids this overcounting by inserting the effective mass \(M^2\) into the first derivative of the effective potential. At one-loop order, the resummed effective potential is obtained by integrating over the background field,
\begin{align}
    V_{\rm eff}^{\rm PD}(\phi,T) = V_0(\phi) + \int_0^\phi \mathrm{d}\phi^\prime~\left(\frac{\partial V_{\rm CW} (M^2)}{\partial M^2} + \frac{\partial V_{T} (M^2,T)}{\partial M^2} \right)\frac{\partial m^2}{\partial \phi^\prime} ,
\end{align}
where the lower integration limit is set to zero, as the effective potential is conventionally normalized to vanish at the origin.

The PD prescription directly resums the daisy and superdaisy diagrams without relying on the high-temperature approximation, making it applicable consistently over the full-temperature regime. However, the gap equation in the PD prescription includes self-energy diagrams with explicit dependence on the external momentum,
\begin{align}
\adjustbox{valign=c, width=0.12\textwidth}{\includegraphics{self-energy/1PI.pdf}}
=
\adjustbox{valign=c, width=0.12\textwidth}{\includegraphics{self-energy/propogator.pdf}}
+
\adjustbox{valign=c, width=0.15\textwidth}{\includegraphics{self-energy/1loopprop.pdf}}
+
\adjustbox{valign=c, width=0.15\textwidth}{\includegraphics{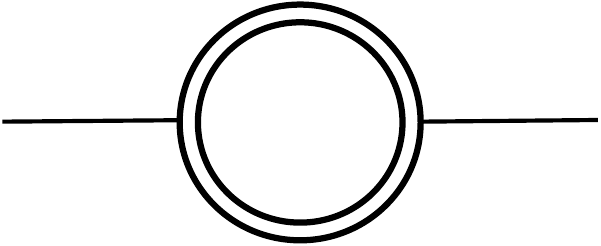}}.
\end{align}
Naively taking the zero-external-momentum limit for all Matsubara frequencies is therefore inconsistent. Moreover, even for the zero mode, this approximation is valid only for daisy diagrams and breaks down when superdaisy contributions are included. The \(k\)-dependent self-energy can also develop an IR divergent term of the form \(\lambda \phi^2 T/M\) in the \(k\to0\) limit, which prevents the gap equation from being solved reliably near \(M^2=0\).

\section{Calculation Details Relevant to the Implementation of the IPD}
Firstly, to determine the phase transition, we require a smooth function of the effective mass, \(M^2(\phi,T)\), with respect to the background field and temperature. Since the gap equation generally forms a system of nonlinear equations that cannot be solved analytically, we solve Eq.~\eqref{mpdgeq} numerically on a \((\phi,T)\) grid and interpolate the resulting values. For the numerical solution of the gap equation, we employ an iterative procedure,
\begin{align}
M^2_{(\ell+1)}
=(1-\xi)M^2_{(\ell)}
+\xi\left[
m^2+\Pi_4\left(M^2_{(\ell)}\right)+\delta m^2
\right].
\end{align}
Here, the initial value \(M^2_{(0)}\) is taken to be the corresponding tree-level value, while the damping parameter \(\xi\) is introduced to improve the convergence of the iteration.

Another key numerical calculation involves the resummed one-loop tadpole diagrams. The sum-integral corresponding to the first term in Eq.~\eqref{eq:1looptadpole}, which only contains the non-zero-mode contributions, can be evaluated as
\begin{align}
\frac{1}{2}g_i\sint_K^\prime \frac{1}{K^2+M_i^2}
=&\ \frac{1}{2}g_i\sint_K\frac{1}{K^2+M_i^2}
-\frac{1}{2}g_iT\int\frac{\mathrm{d}^3k}{(2\pi)^3}
\frac{1}{k^2+M_i^2}
\nonumber\\
=&\ \frac{\partial (V_{\rm CW}+V_T)}{\partial m_i^2}\Bigg|_{m^2_i\to M^2_i}
+\frac{g_iM_iT}{8\pi}.
\end{align}
For brevity, the details of the regularization and subtraction of ultraviolet (UV) divergences are omitted here. For the second term in Eq.~\eqref{eq:1looptadpole}, the momentum integral exhibits a linear UV divergence, which can be removed by subtracting the UV divergent contribution. This gives
\begin{align}
    \int\frac{\mathrm{d}^3k}{(2\pi)^3} \frac{1}{k^2 + M^2_i +\Pi_{i,3}(k)}\to -\frac{1}{2\pi^2}\int_0^\infty\mathrm{d}k~ \frac{M^2_i + \Pi_{i,3}^{\rm heavy}(0)+ \Pi^{\rm light}_{i,3}(k)}{k^2 + M^2_i + \Pi_{i,3}^{\rm heavy}(0) + \Pi_{i,3}^{\rm light}(k)},\label{eq:zeromodeint}
\end{align}
where the momentum-dependent self-energy from a light degree of freedom takes a commonly encountered form
\begin{align}
\Pi^{\rm light}_{i,3}(k)
=& - \frac{1}{2}\left(\frac{\partial m^2_j}{\partial \phi}\right)^2T\int\frac{\mathrm{d}^3p}{(2\pi)^3}~\frac{1}{(p^2+M^2_j )((p+k)^2+M^2_j)}\nonumber\\
=& -\frac{1}{2}\left(\frac{\partial m^2_j}{\partial \phi}\right)^2\frac{ T}{4\pi } \frac{  \arctan\left(\frac{k}{2M_j}\right)}{k}.\label{eq:mdependence}
\end{align}
The integral in Eq.~\eqref{eq:zeromodeint} can then be evaluated numerically. However, a subtlety arises when the denominator of the integrand  has a simple zero, in which case the Cauchy principal value of the integral must be handled carefully in the numerical evaluation. For this case, the zero \(k_0\) can only be found numerically, satisfying
\begin{align}
k_0^2 + M^2_i + \Pi^{\rm heavy}_{i,3}(0)+\Pi^{\rm light}_{i,3}(k_0) \simeq 0.
\end{align}
We rewrite the integrand as
\begin{align}
\frac{f(k)}{g(k)} = \frac{(k-k_0)f(k)/(g(k) - g(k_0))}{k-k_0} ,
\end{align}
where we define
\begin{align}
f(k) \equiv M^2_i + \Pi^{\rm heavy}_{i,3}(0)+\Pi^{\rm light}_{i,3}(k),\quad g(k) \equiv k^2 + M^2_i + \Pi^{\rm heavy}_{i,3}(0)+\Pi^{\rm light}_{i,3}(k).
\end{align}
To render the integration variable dimensionless, we perform the change of variables \(k = \frac{Tt}{1-t}\), which gives \(k_0 = \frac{Tt_0}{1-t_0}\). Then the integral becomes
\begin{align}
    \int_0^\infty\mathrm{d}k~\frac{f(k)}{g(k)} = T\int_0^1\mathrm{d}t~\frac{G(t)}{t-t_0},\label{eq:CPV}
\end{align}
where
\begin{align}
    G(t) = \left\{\begin{aligned}
       & \frac{(t-t_0)}{(1-t)^2}f\left(\frac{Tt}{1-t}\right)\Big/\left[g\left(\frac{Tt}{1-t}\right) - g\left(\frac{Tt_0}{1-t_0}\right)\right]&,\quad |t-t_0|\geq\delta t,\\
        &\frac{f\left(\frac{Tt_0}{1-t_0}\right)}{Tg^\prime\left(\frac{Tt_0}{1-t_0}\right)} &,\quad |t-t_0|<\delta t.
    \end{aligned}\right.
\end{align}
Here, in the small neighborhood of \(t_0\) with \(|t-t_0|<\delta t\), we employ the Taylor expansion \(g(k)=g(k_0) + g'(k_0)(k-k_0) + \cdots\) and subsequently take the limit \(t\to t_0\). And the singularity equation becomes
\begin{align}
    \frac{t^2_0}{(1-t_0)^2} + \frac{M^2_i+ \Pi^{\rm heavy}_{i,3}(0)+\Pi^{\rm light}_{i,3}\left(\frac{Tt_0}{1-t_0}\right)}{T^2} = 0. 
\end{align}
The Cauchy principal value of the integral is evaluated as
\begin{align}
    \int_0^1\mathrm{d}t~\frac{G(t)}{t-t_0} = &\int_0^1\mathrm{d}t~\left(\frac{G(t_0)}{t-t_0} + \frac{G(t)-G(t_0)}{t-t_0}\right)\nonumber\\
    =& G(t_0)\ln\left(\frac{1-t_0}{t_0}\right) + \int_0^1\mathrm{d}t~ \frac{G(t)-G(t_0)}{t-t_0}.
\end{align}
For the singularity-free case, direct numerical integration is feasible, so there is no need to cast the integrand into the singular form \(G(t)/(t-t_0)\).

When the momentum-dependent self-energy Eq.~\eqref{eq:mdependence} is analytically continued to a negative squared mass, \(M^2_j<0\), it gives
\begin{align}
        \Re e\left[\Pi^{\rm light}_{i,3}(k) \right] =&  -\frac{1}{2}\left(\frac{\partial m^2_j}{\partial \phi}\right)^2\frac{ T}{4\pi}\times\left\{\begin{aligned}
       0, &\quad k< 2|M_j|\\
       \frac{\pi}{2k}, &\quad k\geq 2|M_j|
   \end{aligned}\right.. \label{eq:mdepself_real}
\end{align}
This piecewise expression is discontinuous at \(k=2|M_j|\). In particular, when the denominator of the integrand in Eq.~\eqref{eq:zeromodeint} vanishes at \(k=2|M_j|\), the momentum integral develops a logarithmic divergence. To avoid this divergence, the imaginary part of the momentum-dependent self-energy is retained, and the real part of the full integrand is taken before numerical integration. 
This gives
\begin{align}
    &\int_0^\infty\mathrm{d}k~ \Re e\left[\frac{\overline{M}^2_i + \Pi^{\rm light}_{i,3}(k)}{k^2 + \overline{M}^2_i + \Pi_{i,3}^{\rm light}(k)}\right] \nonumber\\
    =& \int_0^\infty\mathrm{d}k~ \frac{\left(\overline{M}^2_i + \Re e\left[\Pi^{\rm light}_{i,3}(k)\right]\right)\left(k^2 + \overline{M}^2_i + \Re e\left[\Pi_{i,3}^{\rm light}(k)\right]\right) + \left(\Im m\left[\Pi_{i,3}^{\rm light}(k)\right]\right)^2}{\left(k^2 + \overline{M}^2_i + \Re e\left[\Pi_{i,3}^{\rm light}(k)\right]\right)^2 + \left(\Im m\left[\Pi_{i,3}^{\rm light}(k)\right]\right)^2},
\end{align}
where
\begin{align}
    \overline{M}_i^2 \equiv M_i^2 + \Pi_{i,3}^{\rm heavy}(0), \quad \Im m\left[\Pi^{\rm light}_{i,3}(k)\right] =  -\frac{1}{2}\left(\frac{\partial m^2_j}{\partial \phi}\right)^2\frac{ T}{4\pi}\times \frac{1}{2k} \log\left|\frac{k+2|M_j|}{k-2|M_j|}  \right|.\label{eq:mdepself_imag}
\end{align}
We can note that the imaginary part of the momentum-dependent self-energy of a light degree of freedom diverges at \(k=2|M|\) for \(M^2<0\). Interpreting this imaginary part as the decay width of an unstable field configuration, as discussed above, would therefore imply an infinite decay width, which is clearly unphysical. When \(|M^2|\) is relatively large, the imaginary part has only a small effect on the integral, consistent with our treatment of heavy degrees of freedom, for which the imaginary contribution is neglected. By contrast, when \(M^2\simeq0\), the imaginary part dominates the integral. Therefore, this singular imaginary contribution should be retained.

\section{Calculations of the effective potential in IDM}
The field-dependent tree-level masses are given by
\begin{align}
    m_h^2(\phi) =&\ \mu_1^2 + \frac{3}{2}\lambda_1\phi^2,\\
    m^2_G(\phi)=&\ m_{G^+}^2(\phi)= m_{G^0}^2(\phi) =  \mu_1^2 + \frac{1}{2}\lambda_1\phi^2,\\
    m_A^2(\phi) =&\ \mu_2^2 + \frac{1}{2}(\lambda_3 + \lambda_4 - \lambda_5)\phi^2 = \mu_2^2 + \frac{1}{2}\lambda_-\phi^2,\\
    m_H^2(\phi) = &\ \mu_2^2 + \frac{1}{2}(\lambda_3 + \lambda_4 + \lambda_5)\phi^2 = \mu_2^2 + \frac{1}{2}\lambda_+\phi^2,\\
    m_{H^{\pm}}^2(\phi) =&\  \mu_2^2 + \frac{1}{2}\lambda_3\phi^2, \\
    m^2_t = &\ \frac{y_t^2\phi^2}{2}.
\end{align}
For the fermionic sector, only the top-quark Yukawa coupling is retained, as the contributions from the other fermions are negligible. The field-dependent mass matrix of the gauge bosons $B_\mu$ and $W^a_\mu$ is
\begin{align}
    \mathcal{M}^2(\phi) = & \begin{pmatrix}
        g^2\phi^2/4 & 0 & 0&0\\
        0& g^2\phi^2/4 & 0 &0\\
        0& 0& g^2\phi^2/4& -g^\prime g \phi^2/4\\
        0& 0& -g^\prime g \phi^2/4& g^{\prime 2}\phi^2/4
    \end{pmatrix},
\end{align}
where \(g\) and \(g^\prime\) denote the gauge couplings associated with \(SU(2)_L\) and \(U(1)_Y\), respectively.
Diagonalizing this matrix yields the gauge boson mass eigenstates \(W^{\pm}_\mu\), \(Z_\mu\), and \(A_\mu\), with the corresponding field-dependent masses 
\begin{align}
    m_W^2(\phi)=\frac{g^2\phi^2}{4},\quad
m_Z^2(\phi)=\frac{(g^2+g^{\prime 2})\phi^2}{4},\quad
m_\gamma^2(\phi)=0.
\end{align}

\subsection{One-loop effective potential without resummation}\label{sec:1loopepnore}

\begin{figure}[t]
    \centering
    \includegraphics[width=0.3\linewidth]{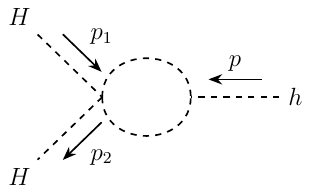}
    \caption{The three-point interaction vertex for the SM Higgs and inert scalar \(H\).}
    \label{fig:hHH}
\end{figure}

We start with the zero temperature effective potential. The tree-level potential is given by
\begin{align}
V_0(\phi)=\frac{1}{2}\mu_1^2\phi^2+\frac{1}{8}\lambda_1\phi^4 .
\end{align}
We employ the \(\overline{\text{MS}}\) scheme throughout this work, with the UV counterterms presented in Sec.~\ref{app:UV-ct}. The one-loop zero temperature effective potential is then given by Eq.~\eqref{eq:CWpotential}. To relate the parameters in the potential to the physical observables, we employ the tree-level relations \(\hat m_h^2=m_h^2(v_0),\
v_0^2=-2\mu_1^2/\lambda_1.\)
This requires us to introduce finite counterterms 
\begin{align}
    V_{\rm CT} = \frac{1}{2}\delta\bar{\mu}_{1}^2 \phi^2 + \frac{1}{8}\delta\bar{\lambda}_{1}\phi^4
\end{align}
to preserve the vacuum location. These counterterms can be determined by imposing the on-shell conditions
\begin{align}
    \frac{\partial(V_{\rm CW} + V_{\rm CT})}{\partial\phi}\Big|_{\phi=v_0} =&\ 0,\label{eq:onshell1}\\
    \frac{\partial^2(V_{\rm CW} + V_{\rm CT})}{\partial\phi^2}\Big|_{\phi=v_0} =&\ - \left[\Sigma_h(\hat{m}_h) - \hat{m}_h^2\delta_Z  \right] + \Sigma_h(0) ,
\end{align}
where \(\Sigma_h(p)\) denotes the Higgs self-energy at zero temperature, \(\hat m_h\) is the physical Higgs pole mass, and \(\delta_Z\) is the wave-function renormalization counterterm, defined as
\begin{equation}
\frac{\partial\Sigma_h(p)}{\partial p^2}\Big|_{p^2=\hat m_h^2}
=\delta_Z .
\end{equation}
The second condition accounts for the difference between the definition of the physical Higgs mass at \(p^2=\hat m_h^2\) and that of the effective potential at zero external momentum. This treatment also avoids the IR divergence associated with the Goldstone contribution~\cite{Delaunay:2007wb}. In particular, \(\Sigma_h(0)\) contains an IR divergence, which is exactly reproduced by the second derivative of the Coleman-Weinberg potential. Consequently, the counterterm contribution satisfies
\begin{align}
\frac{\partial^2 V_{\rm CT}}{\partial\phi^2}\Big|_{\phi=v_0}
=
-\left[\Sigma_h(\hat m_h)-\hat m_h^2\delta_Z\right].
\end{align}
Combining this result with Eq.~\eqref{eq:onshell1}, we can obtain \(\delta \bar{\mu}_1\) and \(\delta\bar{\lambda}_{1}\). For the inert scalar sector, the renormalization conditions are chosen as
\begin{align}
    \delta \bar{\mu}_2^2 + \frac{1}{2}\delta\bar{\lambda}_-v_0^2 =& -\Sigma_A(0) \big|_{\phi = v_0},\\
    \delta\bar{\mu}_2^2 + \frac{1}{2}\delta\bar{\lambda}_+v_0^2 =& -\Sigma_H(0) \big|_{\phi = v_0},\\
     \delta\bar{\mu}_2^2 + \frac{1}{2}\delta\bar{\lambda}_3v_0^2 =& -\Sigma_{H^\pm}(0) \big|_{\phi = v_0},
\end{align}
where \(\delta \bar{\lambda}_- = \delta \bar{\lambda}_3 + \delta \bar{\lambda}_4- \delta \bar{\lambda}_5\) and \(\delta \bar{\lambda}_+ = \delta \bar{\lambda}_3 + \delta \bar{\lambda}_4+\delta \bar{\lambda}_5\). We choose $\hat{m}^2_A = m_A^2(v_0),\ \hat{m}_H^2 = m_H^2(v_0),\ \hat{m}^2_{H^\pm}=m^2_{H^\pm}(v_0),\ \lambda_2$ and $\lambda_{L} = (\lambda_3+\lambda_4+\lambda_5)/2$ as the physical input parameters. The fourth renormalization condition is then given by
\begin{align}
    \delta\bar{\lambda}_+v_0 + \Gamma_{hHH}(p = \hat{m}_h) = 0,
\end{align}
where \(\Gamma_{hHH}\) denotes the one-loop correction to the \(h\)-\(H\)-\(H\) vertex, as shown in Fig.~\ref{fig:hHH}.
The contributions classified by the species of loop particles read
\begin{align}
    \Gamma_{hHH}^{h}(p) =&  -\frac{3}{2}\lambda_+ \lambda_1 v_0 \int_0^1 \mathrm{d}x~I_L(m_{h,*}^2) ,\\
    \Gamma_{hHH}^{G}(p) =&  -\left(\frac{1}{2}\lambda_- + \lambda_3\right) \lambda_1 v_0 \int_0^1\mathrm{d}x~ I_L(m_{G,*}^2) ,\\
    \Gamma_{hHH}^{A}(p) =&  -\frac{1}{2}\lambda_2 \lambda_- v_0 \int_0^1\mathrm{d}x~I_L(m_{A,*}^2) ,\\
    \Gamma_{hHH}^{H}(p) =&  -\frac{3}{2}\lambda_2 \lambda_+ v_0 \int_0^1\mathrm{d}x~I_L(m_{H,*}^2)  ,\\
    \Gamma_{hHH}^{H^\pm}(p) =&  -\lambda_2 \lambda_3 v_0\int_0^1\mathrm{d}x~I_L(m_{H^\pm,*}^2)  ,\\
    \Gamma_{hHH}^{Z}(p) =&  -\frac{3}{8}(gc_w + g^\prime s_w)^4 v_0 \int_0^1\mathrm{d}x~I_L(m_{Z,*}^2) , \\
    \Gamma_{hHH}^{W}(p) =&  -\frac{3}{4}g^4 v_0 \int_0^1\mathrm{d}x~I_L(m_{W,*}^2) ,
\end{align}
where $m^2_{i,*} \equiv m^2_i - x(1-x)p^2$. The loop integral is given by
\begin{align}
    I_L(m^2) = -\frac{1}{16\pi^2}\left(\log\frac{m^2}{\mu_R^2}+c\right),
\end{align}
where \(c=0\) for scalars and fermions, while \(c=2/3\) for vectors.
These lead to an additional finite mass counterterm \(\delta m^2\) in the one-loop mass corrections of the scalars. For the gauge bosons, we adopt the \(\overline{\mathrm{MS}}\) results of the electroweak SM, and therefore no additional finite counterterms are introduced.

According to the particle species running in the loop, the one-loop vertex correction receives the following contributions:
\begin{align}
    \Gamma_{hHH}^{h} =&  -\frac{3}{2}\lambda_+ \lambda_1 v_0 \int_0^1 \mathrm{d}x~I_2^b(m_{h,*}^2, 0) ,\\
    \Gamma_{hHH}^{G} =&  -\left(\frac{1}{2}\lambda_- + \lambda_3\right) \lambda_1 v_0 \int_0^1\mathrm{d}x~ I_2^b(m_{G,*}^2, 0) ,\\
    \Gamma_{hHH}^{A} =&  -\frac{1}{2}\lambda_2 \lambda_- v_0 \int_0^1\mathrm{d}x~I_2^b(m_{A,*}^2, 0) ,\\
    \Gamma_{hHH}^{H} =&  -\frac{3}{2}\lambda_2 \lambda_+ v_0 \int_0^1\mathrm{d}x~I_2^b(m_{H,*}^2, 0)  ,\\
    \Gamma_{hHH}^{H^\pm} =&  -\lambda_2 \lambda_3 v_0\int_0^1\mathrm{d}x~I_2^b(m_{H^\pm,*}^2, 0)  ,\\
    \Gamma_{hHH}^{Z} =&  -\frac{3-2\epsilon}{8}(gc_w + g^\prime s_w)^4 v_0 \int_0^1\mathrm{d}x~I_2^b(m_{Z,*}^2, 0) , \\
    \Gamma_{hHH}^{W} =&  -\frac{3-2\epsilon}{4}g^4 v_0 \int_0^1\mathrm{d}x~I_2^b(m_{W,*}^2, 0) ,
\end{align}
where \(\lambda_- = \lambda_3 + \lambda_4 - \lambda_5\), \(\lambda_+ = \lambda_3 + \lambda_4 + \lambda_5\) and \(m^2_{i,*}\equiv m_i^2-x(1-x)p^2\).

We next extend the effective potential to finite temperature by including the one-loop thermal contribution. It is useful to introduce the following functions,
\begin{align}
I^{b/f}_0(m^2,T) \equiv& -\sint_K
\log\left(\omega_n^2+\mathbf{k}^2+m^2\right),\label{eq:Ibf0}\\
I^{b/f}_j(m^2,T) \equiv&\ \sint_K
\frac{1}{(\omega_n^2+\mathbf{k}^2+m^2)^j},\quad j= 1,2,3,...,\label{eq:Ijfi}
\end{align}
which satisfy the recursion relation
\begin{equation}
     I^{b/f}_j(m^2,T) = \frac{1}{(j-1)!}~(-1)^{j}~\frac{\mathrm{d}^jI^{b/f}_0(m^2,T) }{\mathrm{d}(m^2)^j}.
\end{equation}
The superscripts ``\(b\)'' and ``\(f\)'' denote the bosonic and fermionic cases, respectively, with the corresponding Matsubara frequencies \(\omega_n=2n\pi T\) and \(\omega_n=(2n+1)\pi T\). The function \(I^{b/f}_0\) can be further separated into the zero temperature and thermal parts,
\begin{align}
    I^{b/f}_0(m^2,T) = -\underbrace{\frac{m^4}{32\pi^2} \left(-C_{\rm UV} + \log \frac{m^2}{\mu_R^2} - \frac{3}{2} \right)}_{\mathrm{zero-}T~\mathrm{part}} -  \underbrace{\frac{ T^4}{\pi^2}J_{b/f}\left(\frac{m^2}{T^2} \right)}_{\mathrm{finite-}T~\mathrm{part}},
\end{align}
where \(C_{\rm UV}=1/\epsilon+\log(4\pi)-\gamma_E\) denotes the UV subtraction term in the \(\overline{\mathrm{MS}}\) scheme, with dimensional regularization performed in \(4-2\epsilon\) spacetime dimensions. The one-loop finite-temperature effective potential is then given by
\begin{align}
    V_1(\phi,T) =& -\frac{1}{2}\sum_{i=\mathrm{bosons}} g_i\left[I^{b}_0(m_i^2,T)  - V_{\rm UV}^i(m_i^2) \right]+ 6\left[I^{f}_0(m_t^2,T) - V_{\rm UV}^t(m_t^2)\right] + V_{\rm CT}(\phi) \nonumber\\
    &- \sum_{i=W_T,Z_T,\gamma_T}\epsilon V_{\rm UV}^i(m_i^2) ,
\end{align}
where the UV counterterm is 
\begin{align}
    V_{\rm UV}^i  = \frac{ m_i^4(\phi)}{32\pi^2} C_{\rm UV} .
\end{align}
The appearance of the \(\epsilon\) term originates from the identity \(g_{\mu\nu}g^{\mu\nu}=4-2\epsilon\) in dimensional regularization, with the limit \(\epsilon\to0\) taken at the end of the calculation.

\subsection{One-loop effective potential with resummation}
We next consider the resummation effects. In the Parwani and AE schemes, which restore the high-temperature ring-improved expansion, the bosonic thermal masses are replaced as \(m_i^2\to M_i^2=m_i^2+\Pi_i\), where
\begin{align}
    &\Pi_h = \Pi_G = \frac{T^2}{12} \left(\frac{9 g^2+3 g^{\prime 2}}{4}+ 3 \lambda_1 +2\lambda_3+\lambda_4+3 y_t^2\right),\\
     &\Pi_A = \Pi_H = \Pi_{H^\pm}=\frac{T^2}{12} \left(\frac{9 g^2+3 g^{\prime 2}}{4}+ 3 \lambda_2 +2\lambda_3+\lambda_4\right),\\
    & \Pi_{L} = \begin{pmatrix}
      2g^2T & 0& 0& 0 \\
       0&   2g^2 T^2 & 0&0\\
       0& 0& 2g^2 T^2 & 0\\
        0& 0& 0 & 2g^{\prime 2} T^2
     \end{pmatrix},\quad \Pi_{T} = \mathbf{0}_{4\times 4}.
\end{align}
For the longitudinal components of the gauge bosons, the physical masses are obtained from the eigenvalues of the matrix \(\mathcal{M}^2+\Pi_L\). The one-loop resummed effective potentials in the Parwani and AE schemes are then given by
 \begin{align}
 V_{1,R}^{\rm Par}(\phi,T) =& -\frac{1}{2}\sum_{i=\mathrm{bosons}} g_i\left[I^{b}_0(M_i^2,T)  - V_{\rm UV}^i(M_i^2) \right]+ 6\left[I^{f}_0(m_t^2,T) - V_{\rm UV}^t(m_t^2)\right] + V_{\rm CT}(\phi) \nonumber\\
 &- \sum_{i=W_T,Z_T,\gamma_T}\epsilon V_{\rm UV}^i(m_i^2),\\
     V_{1,R}^{\rm AE}(\phi,T) =&\  V_1(\phi,T) - \frac{1}{2}\sum_{i=\mathrm{bosons}}g_i \Big( I_{0,n=0}^b(M_i^2,T) - I_{0,n=0}^b(m_i^2,T) \Big),
 \end{align}
where 
\begin{align}
    I_{0,n=0}^b(m^2,T) = -T \int\frac{\mathrm{d}^3 k}{(2\pi)^2}\log (k^2+m^2) = \frac{m^3T}{6\pi} .
\end{align}

For the DR scheme, we use the package \texttt{DRalgo}~\cite{Ekstedt:2022bff,Bernardo:2026nyq,Fonseca:2020vke} to perform the DR and compute the effective potential. In this work, only the hard scales, corresponding to all non-zero Matsubara modes, are integrated out. The couplings are matched at one-loop order and the masses are matched at two-loop order, while the resulting effective potential is evaluated only at one-loop order, \(V_{3d,\rm eff}(\phi_{3d},T)=V_{3d,0}+V_{3d,1}.\) 
For comparison with the four-dimensional effective potentials, we convert the three-dimensional effective potential into a four-dimensional form, \(V_{4d,\rm eff}=T V_{3d,\rm eff},\) and employ the approximate relation \(\phi_{4d}\simeq \sqrt{T}\phi_{3d}.\) In addition, we note that within the DR scheme, the input parameters are \(\mu_R\)-dependent \(\overline{\mathrm{MS}}\) renormalized parameters, whereas the remaining four-dimensional schemes employ on-shell renormalized parameters. Their relation reads
\begin{align}
   \lambda_{MS}(\mu_R) + \delta \lambda_{MS}(\mu_R) = \lambda_{OS} + \delta\lambda_{OS}.
\end{align}
Here \(\delta \lambda_{MS}(\mu_R) \) cancels the \(C_{\mathrm{UV}}\) piece, so that
\begin{align}
    \lambda_{MS}(\mu_R) = \lambda_{OS} + \delta\bar{\lambda}_{OS},
\end{align}
where \(\delta\bar{\lambda}_{\mathrm{OS}}\) is precisely the finite counterterm for zero temperature renormalization discussed in Sec.~\ref{sec:1loopepnore}.

For the PD and IPD resummation prescriptions, the gap equations need to be solved numerically. The transverse and longitudinal components of the gauge bosons are treated as independent degrees of freedom. Therefore, the gap equation is, in principle, an 11-dimensional system of nonlinear equations. However, in the numerical calculation, the self-energy contributions of the transverse gauge bosons are found to be negligible. We therefore retain their tree-level masses, reducing the gap equation to an 8-dimensional system. This treatment also improves the numerical stability of the solution. Since the \(W^3\) and \(B\) gauge bosons are mixed, an additional diagonalization step is required for the gauge boson sector of the gap equation,
\begin{align}
    M_{\rm GB}^2 = V_{\rm EW}\left(\mathcal{M}^2 + \Pi_L(M_{\rm GB}^2)\right)V_{\rm EW}^{\rm T}.
\end{align}
Here, the orthogonal matrix
\begin{align}
    V_{\rm EW} = \begin{pmatrix}
    1\ \  & 0\ \ & 0&0\\
    0\ \ &1\ \ &0&0\\
    0\ \ &0\ \ &    \tilde{c}_w & -\tilde{s}_w\\
     0\ \ &0\ \ &   \tilde{s}_w & \tilde{c}_w
    \end{pmatrix},\quad \tilde{c}_w \equiv\cos\tilde{\theta},\quad \tilde{s}_w\equiv\sin\tilde{\theta},
\end{align}
is parameterized by the effective mixing angle \(\tilde{\theta}\). During the numerical iteration, the effective mixing angle is also updated self-consistently. The complete expressions for the one-loop self-energies of all bosonic fields are provided in Sec.~\ref{sec:self-energy}. 

There is a subtlety in solving the gap equation regarding the treatment of finite counterterms. For the PD scheme, the full self-energy is incorporated into the gap equation, so including the entire finite counterterm therein is self-consistent and automatically satisfies the zero temperature on-shell mass conditions. This is, however, not the case for the IPD scheme: only the momentum-independent part of the self-energy enters the gap equation in this scheme, and one cannot simply insert all finite counterterms into it. We note that the momentum-dependent self-energy \(\Pi_3\) contributes solely to \(\delta\bar{\lambda}_i\). Accordingly, we decompose the finite counterterms as
\(\delta\bar{\lambda}_i = \delta\bar{\lambda}_{i,3} + \delta\bar{\lambda}_{i,4}.\)
We then impose the on-shell renormalization conditions
\begin{align}
     \Pi_{h,4}(\hat{m}^2_i)\Big|_{\phi = v_0,T=0} + \delta \bar{\mu}^2_1 + \frac{3}{2}\delta\bar{\lambda}_{1,4}v_0^2  & =0,\\
   \Pi_{A,4}(\hat{m}^2_i)\Big|_{\phi = v_0,T=0} + \delta \bar{\mu}^2_2 + \frac{1}{2}\delta\bar{\lambda}_{-,4}v_0^2  &= 0 ,\\
    \Pi_{H,4}(\hat{m}^2_i)\Big|_{\phi = v_0,T=0} + \delta \bar{\mu}^2_2 + \frac{1}{2}\delta\bar{\lambda}_{+,4}v_0^2  &=0,\\
    \Pi_{H^\pm,4}(\hat{m}^2_i)\Big|_{\phi = v_0,T=0} + \delta \bar{\mu}^2_2 + \frac{1}{2}\delta\bar{\lambda}_{3,4}v_0^2 & =0.
\end{align}

We solve the gap equations on the two-dimensional grid of interest, \(\phi \in [0,400]~\mathrm{GeV}\) and \(T\in [50,300]~\mathrm{GeV}\). In the PD scheme, the solutions become unreliable in the regions where \(M_h^2(\phi,T)\sim 0\) or \(M_G^2(\phi,T)\sim 0\). In these regions, we estimate the solutions by fitting the surrounding data points and subsequently perform a uniform two-dimensional interpolation. The resulting one-loop resummed effective potential in the PD scheme is given by
\begin{align}
     V_{1,R}^{\mathrm{PD}}(\phi,T) =&\ 6\left[I^f_0(m^2_t, T) - V^t_{\rm UV}(m^2_t)\right] + V_{\rm CT}(\phi) - \sum_{i=W_T,Z_T,\gamma_T}\epsilon V^i_{\rm UV}(m^2_i) \nonumber\\
     &+ \frac{1}{2}\int_0^\phi \mathrm{d}\phi^\prime \sum_{i=\mathrm{bosons}}g_i\left[I^b_1(M_i^2, T)  + \frac{\partial V^i_{\rm UV}(M^2_i)}{\partial M_i^2}\right]\frac{\partial m_i^2}{\partial \phi^\prime}.
\end{align}
In the IPD scheme, only the Higgs and Goldstone bosons are treated as light degrees of freedom. Therefore, only the zero-mode contributions to the Higgs self-energy, \(\Pi_h^{hh,0}\) and \(\Pi_h^{GG,0}\), corresponding to the Higgs and Goldstone boson loops, respectively, need to retain their momentum dependence. The Higgs zero-mode tadpole contribution is given by
\begin{align}
    \partial_\phi V^{n=0}_{h} =&\ \frac{3}{2}\lambda_1\phi T\int\frac{\mathrm{d}^3k}{(2\pi)^3}~ \Re e\left[\frac{1}{k^2 + M^2_h + \Pi^{\mathrm{heavy}}_{h,3}(0) +  \Pi^{hh,0}_h(k) + \Pi^{GG,0}_h(k)}\right]\nonumber\\
    =& \ -\frac{3}{4\pi^2}\lambda_1\phi T\int_0^\infty\mathrm{d}k~\Re e\left[\frac{M^2_h + \Pi^{\mathrm{heavy}}_{h,3}(0) +  \Pi^{hh,0}_h(k) + \Pi^{GG,0}_h(k)}{k^2 + M^2_h + \Pi^{\mathrm{heavy}}_{h,3}(0) +  \Pi^{hh,0}_h(k) + \Pi^{GG,0}_h(k)}\right]
\end{align}
where \(\Pi^{hh,0}_h(k) = - \frac{9}{2}\lambda_1^2\phi^2 I_2^{n=0}(M_{h}^2,k,T)\) and \( \Pi^{GG,0}_h(k) = - \frac{3}{2}\lambda_1^2\phi^2 I_2^{n=0}(M_{G}^2,k,T)\) with
\begin{align}
     I_2^{n=0}(m^2,k,T) =& \int\frac{\mathrm{d}^3p}{(2\pi)^3}~\frac{1}{(p^2+m^2 )((p+k)^2+m^2)} \nonumber\\
        =& \frac{T}{4\pi}~\frac{\arctan\left(\frac{k}{2m}\right)}{k}.
\end{align}
The one-loop resummed effective potential in the IPD scheme is then given by
\begin{align}
    V^{\rm IPD}_{1,R}(\phi,T) =&\ 6\left[I^f_0(m^2_t, T) - V^t_{\rm UV}(m^2_t)\right] + V_{\rm CT}(\phi) - \sum_{i=W_T,Z_T,\gamma_T}\epsilon V^i_{\rm UV}(m^2_i) \nonumber\\
    & + \frac{1}{2}\int^\phi_0\mathrm{d}\phi^\prime \left\{\sum_{i=\mathrm{bosons}}g_i\left[I^b_{1,n\neq 0}(M_i^2, T)  + \frac{\partial V^i_{\rm UV}(M^2_i)}{\partial M_i^2}  \right]\frac{\partial m_i^2}{\partial \phi^\prime}  \right.\nonumber\\
    & \left. + \sum_{i\neq h}g_i I^b_{1,n=0}\left(M^2_i+ \Pi_{i,3}^{\rm heavy}(0),T\right)\frac{\partial m_i^2}{\partial \phi^\prime}  +  \partial_\phi V^{n=0}_{h}  \right\},
\end{align}
where
\begin{align}
    I^b_{1,n=0}(m^2,T) =& - \frac{\partial I^b_{0, n=0}(m^2,T)}{\partial m^2} = - \frac{mT}{4\pi},\\
    I^b_{1,n\neq 0}(m^2,T) =& I^b_1(m^2,T) - I^b_{1,n=0}(m^2,T).
\end{align}

\section{Supplementary Results for Phase Transition Dynamics}
\begin{figure}[t]
    \centering
    \includegraphics[width=\linewidth]{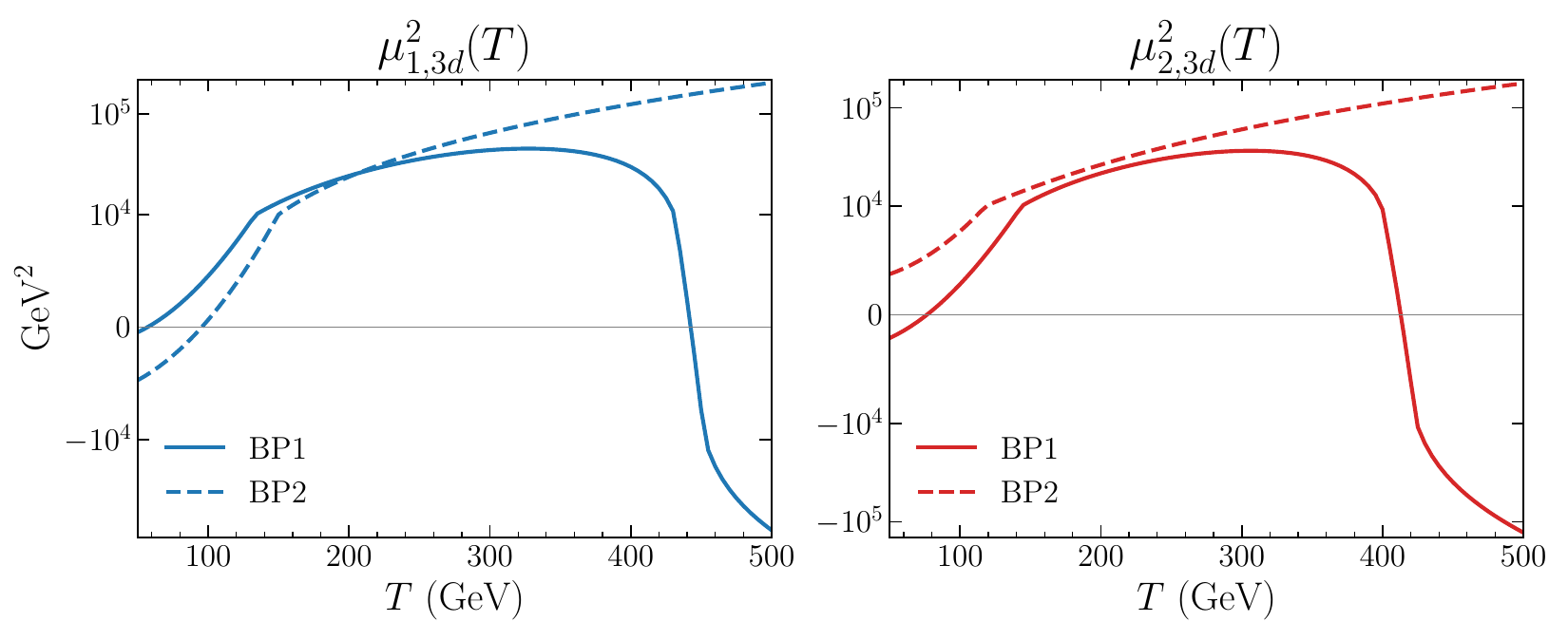}
    \caption{Temperature dependence of the scalar mass-squared parameters in the three-dimensional effective theory. The left and right panels correspond to the first and second doublet, $\mu_{1,3d}^2(T)$ and $\mu_{2,3d}^2(T)$, respectively. Solid lines denote the BP1, while dashed lines show the BP2 for comparison. For BP1, the non-monotonic behavior at high temperatures signals the breakdown of scale separation in the DR scheme. The parameters of BP2 are specified in Table~\ref{tab:BP2_input}.}
    \label{fig:musq_3d}
\end{figure}
\begin{figure}
    \centering
    \includegraphics[width=0.49\linewidth]{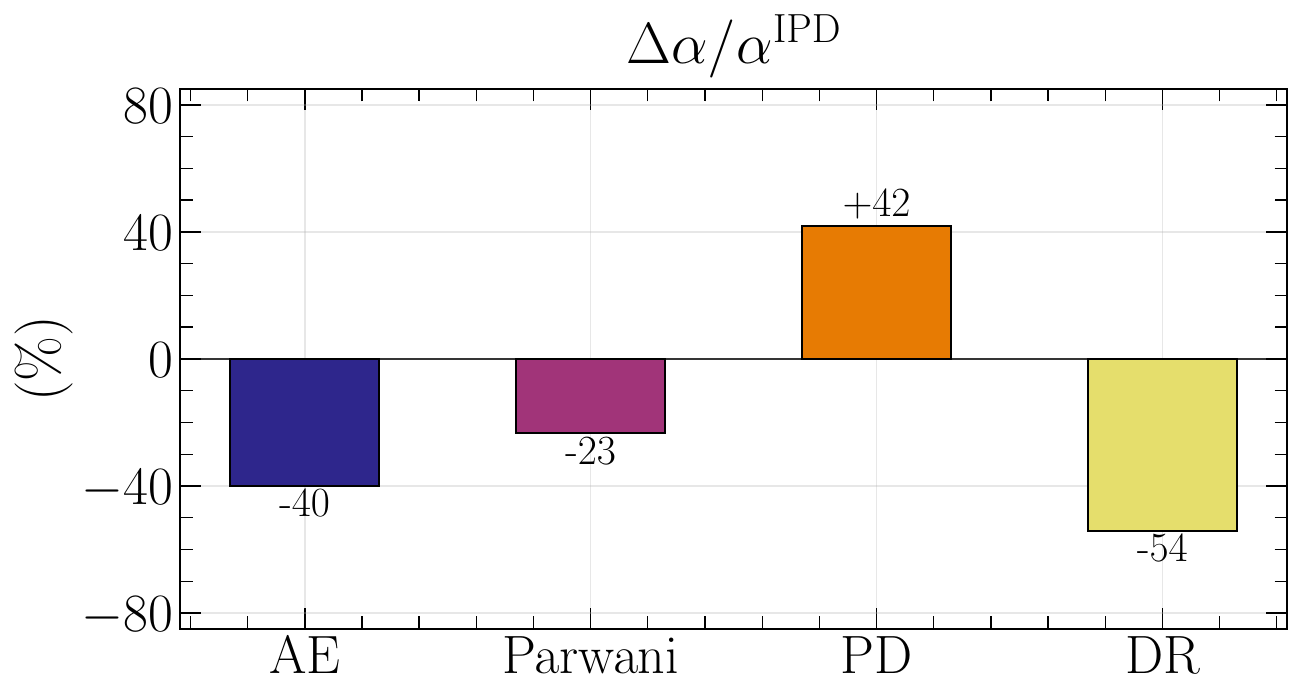}
    \includegraphics[width=0.49\linewidth]{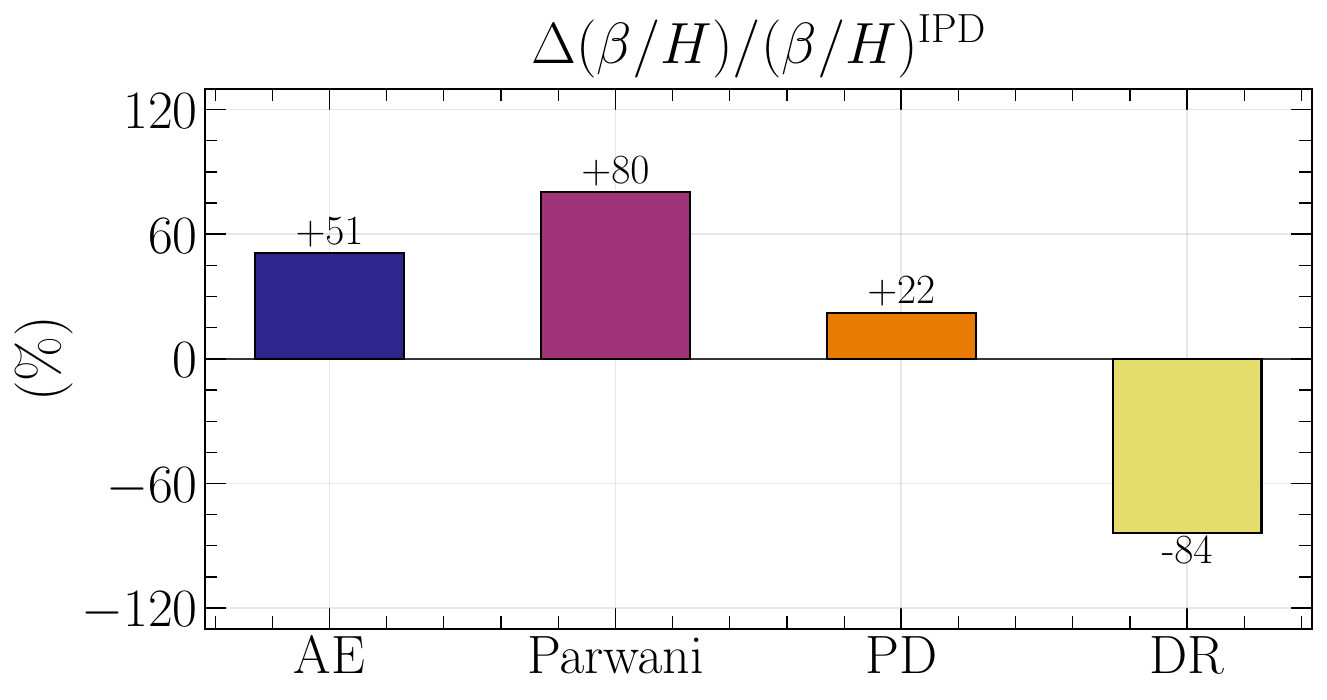}
    \includegraphics[width=0.49\linewidth]{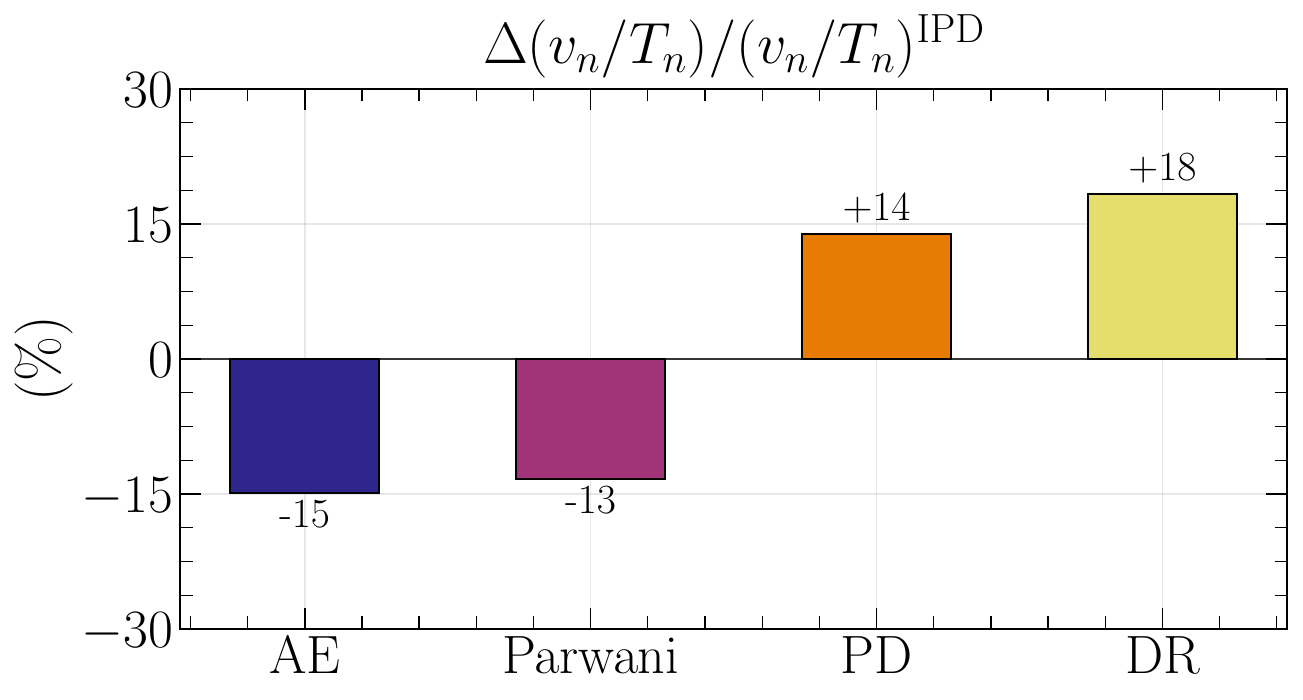}
    \caption{The percentage deviations of FOPT parameters relative to the reference IPD resummation scheme for BP2 with parameters defined in Table~\ref{tab:BP2_input}. The top left panel corresponds to \(\alpha\), the top right panel corresponds to \(\beta/H\) and the bottom panel corresponds to \(v_n/T_n\), respectively. The deviation is defined as \(\Delta X = X^{i} - X^{\rm IPD}\).}
    \label{fig:PTprama}
\end{figure}

In the main text, we find that the high-temperature symmetry non-restoration in the DR scheme occurs at BP1. For further illustration, Fig.~\ref{fig:musq_3d} plots the scalar mass parameter of the three-dimensional effective theory as a function of temperature, with results for the BP2, specified in Table~\ref{tab:BP2_input}, shown for comparison. For BP1, the thermal mass instead decreases at high temperatures, as \(\pi T\) becomes comparable to \(\sqrt{\lambda_3}\,T\) in this parameter set, violating the clear scale separation required by the DR approach. 
In addition, to demonstrate the scheme dependence of phase transition parameters for weak FOPT, we use BP2 (Table~\ref{tab:BP2_input}), as detailed in Table~\ref{tab:scheme_compareBP2}. The scheme sensitivity for weak transitions is visualized in Fig.~\ref{fig:PTprama}, which presents the percentage deviations of \(v_n/T_n\), \(\alpha\), and \(\beta/H\) relative to the IPD reference for BP2 across all considered resummation prescriptions.

\begin{table}[t]
    \caption{Input and derived parameters for BP2.}
    \label{tab:BP2_input}

        \begin{tabular}{|c|c|c|c|}
    \hline
            \makecell{Input\\parameter} & \makecell{Value} & \makecell{Derived\\parameter} & \makecell{Value} \\
            \hline
            $\hat m_h$ & $125.10~\mathrm{GeV}$
            & $\lambda_1$ & $0.26$ \\ \hline
            $v_0$ & $246.22~\mathrm{GeV}$
            & $\mu_1^2$ & $-7825~\mathrm{GeV}^2$ \\ \hline
            $\lambda_2$ & $0.2$
            & $\mu_2^2$ & $4134~\mathrm{GeV}^2$ \\ \hline
            $\lambda_L$ & $0.0015$
            & $\lambda_3$ & $2.83$ \\ \hline
            $\hat m_A$ & $300~\mathrm{GeV}$
            & $\lambda_4$ & $-1.41$ \\ \hline
            $\hat m_{H^\pm}$ & $300~\mathrm{GeV}$
            & $\lambda_5$ & $-1.41$ \\ \hline
            $\hat m_H$ & $65~\mathrm{GeV}$
            & & \\
            \hline
        \end{tabular}
        
\vspace{8pt}
\caption{Comparison of the phase transition parameters obtained using different resummation schemes for BP2, defined by Table~\ref{tab:BP2_input}.}
\label{tab:scheme_compareBP2}
\begin{tabular}{|c|c|c|c|c|c|c|}
\hline
Scheme
&  $T_c(\rm GeV)$ &  $T_n(\rm GeV)$ & $\xrightarrow[\rm type]{\rm Transition}$
 & $v_n/T_n$ &$\alpha$ &\ $\beta/H$ \\
\hline
Parwani & 118.12 & 117.39 & FOPT  & 1.11 & 0.0043 & 36363 \\
\hline
PD & 113.38 & 112.37 & FOPT  & 1.46 & 0.0080 & 24634 \\
\hline
AE & 129.70 & 128.75  & FOPT  & 1.09 & 0.0034 & 30433 \\
\hline
DR& 196.56  & 182.00 & FOPT & 1.52 & 0.0026 & 3262  \\
\hline
IPD & 118.16  &  116.88  & FOPT  & 1.29 & 0.0057 & 20181 \\
\hline
\end{tabular}
\end{table}

\section{Zero Temperature Renormalization}\label{app:UV-ct}
In this section, we present the renormalization of fields and couplings relevant to the computation of the effective potential in the IDM. Bare fields and bare couplings are denoted by the subscript ``\(b\)''. Their relations read
\begin{align}
    &W^a_{\mu,b} \equiv Z^{1/2}_{W} W^a_{\mu} = (1+\delta Z_{W})^{1/2} W^a_\mu,\quad  B_{\mu,b} \equiv Z^{1/2}_{B} B_\mu = (1+\delta Z_{B})^{1/2}B_\mu, \nonumber\\
    &\Phi_b \equiv Z^{1/2}_{\Phi} \Phi = (1+\delta Z_{\Phi})^{1/2} \Phi,\quad  \eta_b \equiv Z^{1/2}_{\eta} \eta = (1+\delta Z_{\eta})^{1/2} \eta,
\end{align}
and
\begin{align}
    & g_b^2 \equiv \mu_R^{2\epsilon} (g^2 + \delta g^2),\quad g_b^{\prime 2} \equiv \mu_R^{2\epsilon} (g'^2 + \delta g'^2) ,\nonumber\\
    & \mu_{1,b}^2 \equiv Z^{-1}_\Phi (\mu_1^2 + \delta \mu_1^2 + \delta \bar{\mu}_1^2),\quad\mu_{2,b}^2 \equiv Z^{-1}_\eta (\mu_2^2 + \delta \mu_2^2 + \delta \bar{\mu}_2^2),\nonumber\\
    & \lambda_{1,b} \equiv Z^{-2}_\phi \mu_R^{2\epsilon}(\lambda_1 + \delta \lambda_1 + \delta \bar{\lambda}_1), \quad \lambda_{2,b} \equiv Z^{-2}_\eta \mu_R^{2\epsilon}(\lambda_2 + \delta \lambda_2 + \delta \bar{\lambda}_2),\nonumber\\
    & \lambda_{3,b} \equiv Z^{-1}_\Phi Z^{-1}_\eta \mu_R^{2\epsilon}(\lambda_3 + \delta \lambda_3 + \delta \bar{\lambda}_3), \quad \lambda_{4,b} \equiv Z^{-1}_\Phi Z^{-1}_\eta \mu_R^{2\epsilon}(\lambda_4 + \delta \lambda_4 + \delta \bar{\lambda}_4),\nonumber\\
    & \lambda_{5,b} \equiv Z^{-1}_\Phi Z^{-1}_\eta \mu_R^{2\epsilon}(\lambda_5 + \delta \lambda_5+ \delta \bar{\lambda}_5).
\end{align}
Therefore, the potential contains a UV-finite counterterm
\begin{align}
\delta \bar{V}=&\delta \bar{\mu}_{1}^{2}\Phi^\dag\Phi+ \delta\bar{\mu}_{2}^{2}\eta^\dag\eta+\frac{1}{2}\delta\bar{\lambda}_{1}(\Phi^\dag\Phi)^{2}+\frac{1}{2}\delta\bar{\lambda}_{2}(\eta^\dag\eta)^{2}+ \delta\bar{\lambda}_{3}(\Phi^\dag\Phi)(\eta^\dag \eta)\nonumber\\
&+ \delta\bar{\lambda}_{4}(\Phi^{\dag}\eta)(\eta^\dag\Phi)+\frac{1}{2}\{\delta\bar{\lambda}_{5}(\Phi^{\dagger}\eta)^{2}+\mathrm{h.c.}\},\label{eq:CTVPhieta}
\end{align}
which is determined by the scalar mass renormalization conditions in Sec.~\ref{sec:1loopepnore}. The UV-divergent counterterms are
{\allowdisplaybreaks
\begin{align}
    \delta Z_W =&\ \frac{g^2C_{\rm UV}}{16\pi^2}\left(\frac{26-N_d}{6} - \frac{4}{3}N_f\right),\\
    \delta Z_B =&\ -\frac{g^{\prime 2}C_{\rm UV}}{96\pi^2}\left(N_d + \frac{40}{3}N_f\right),\\
    \delta Z_\phi =&\ \frac{C_{\rm UV}}{16\pi^2}\left(\frac{9}{4}g^2 + \frac{3}{4}g^{\prime 2} - 3y_t^2\right),\\
    \delta Z_\eta =&\ \frac{C_{\rm UV}}{16\pi^2}\left(\frac{9}{4}g^2 + \frac{3}{4}g^{\prime 2} \right),\\
    \delta g^2 =&\ -\frac{g^4C_{\rm UV}}{16\pi^2}\left(\frac{44-N_d}{6} - \frac{4}{3}N_f \right),\\
    \delta g^{\prime 2} =&\ \frac{g^{\prime 4}C_{\rm UV}}{96\pi^2}\left(N_d + \frac{40}{3}N_f \right),\\
    \delta\mu_1^2 =&\ \frac{C_{\rm UV}}{16\pi^2}\left( 3\lambda_1\mu_1^2 + 2\lambda_3\mu_2^2 + \lambda_4\mu_2^2     \right),\\
     \delta\mu_2^2 =&\ \frac{C_{\rm UV}}{16\pi^2}\left(2\lambda_3 \mu_1^2 + \lambda_4\mu_1^2 + 3\lambda_2\mu_2^2   \right),\\
    \delta\lambda_1 =&\ \frac{C_{\rm UV}}{16\pi^2}\left(6\lambda_1^2  + 2\lambda_3^2+ \lambda_4^2 + 2\lambda_3\lambda_4 + \lambda_5^2 - 6y_t^4 +  \frac{6g^4+ 3(g^2+g^{\prime 2})^2}{8}       \right),\\
    \delta\lambda_2 =&\ \frac{C_{\rm UV}}{16\pi^2}\left(6\lambda_1^2 + 2\lambda_3^2+ \lambda_4^2 + 2\lambda_3\lambda_4 + \lambda_5^2 +  \frac{6g^4+ 3(g^2+g^{\prime 2})^2}{8}    \right),\\
    \delta\lambda_3 =&\ \frac{C_{\rm UV}}{16\pi^2}\left( (\lambda_1 + \lambda_2)(3\lambda_3 + \lambda_4) + 2\lambda_3^2 + \lambda_4^2 + \lambda_5^2 + \frac{6g^4+3(g^{\prime 2}-g^2)^2}{8}\right),\\
    \delta\lambda_4 =&\ \frac{C_{\rm UV}}{16\pi^2}\left(  (\lambda_1+\lambda_2)\lambda_4 + 2\lambda_4^2 + 4\lambda_3\lambda_4 + 4\lambda_5^2 + \frac{3}{2}g^2g^{\prime 2}\right),\\
     \delta\lambda_5 =&\ \frac{\lambda_5C_{\rm UV}}{16\pi^2}\left( \lambda_1+ \lambda_2 + 4\lambda_3 + 6\lambda_4\right),
\end{align}
}
where \(C_{\rm UV}=1/\epsilon+\log(4\pi)-\gamma_E\) denotes the UV subtraction constant in the \(\overline{\mathrm{MS}}\) scheme, \(N_d=2\) is the number of doublets and \(N_f = 3\) is the number of fermion generations.

\section{One-loop Self-energy of Bosonic Degrees of Freedom}\label{sec:self-energy}
\begin{figure}[t]
    \centering
    \subfloat[]{\includegraphics[width=0.25\textwidth]{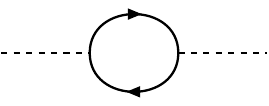}\label{scalar_a}}
    \hspace{1em}
    \subfloat[]{\includegraphics[width=0.25\textwidth]{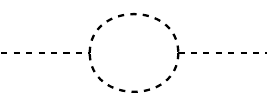}\label{scalar_b}}
    \hspace{1em}
    \subfloat[]{\includegraphics[width=0.25\textwidth]{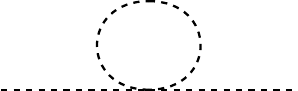}\label{scalar_c}}

    \par\medskip

    \subfloat[]{\includegraphics[width=0.25\textwidth]{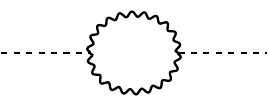}\label{scalar_d}}
    \hspace{1em}
    \subfloat[]{\includegraphics[width=0.25\textwidth]{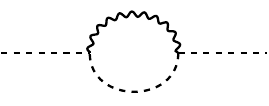}\label{scalar_e}}
    \hspace{1em}
    \subfloat[]{\includegraphics[width=0.25\textwidth]{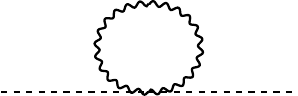}\label{scalar_f}}

    \caption{Scalar boson self-energy diagrams.}
    \label{scalarself}
\end{figure}

\begin{figure}[t]
    \centering
    \subfloat[]{\includegraphics[width=0.25\textwidth]{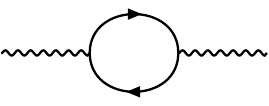}\label{vector_a}}
    \hspace{1em}
    \subfloat[]{\includegraphics[width=0.25\textwidth]{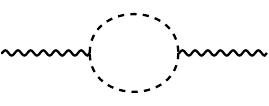}\label{vector_b}}
    \hspace{1em}
    \subfloat[]{\includegraphics[width=0.25\textwidth]{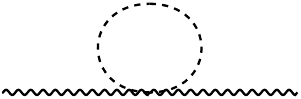}\label{vector_c}}

    \par\medskip

    \subfloat[]{\includegraphics[width=0.25\textwidth]{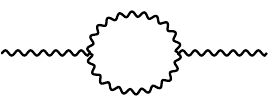}\label{vector_d}}
    \hspace{1em}
    \subfloat[]{\includegraphics[width=0.25\textwidth]{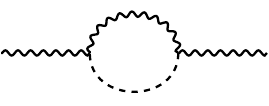}\label{vector_e}}
    \hspace{1em}
    \subfloat[]{\includegraphics[width=0.25\textwidth]{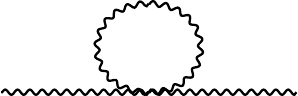}\label{vector_f}}

    \par\medskip

    \subfloat[]{\includegraphics[width=0.25\textwidth]{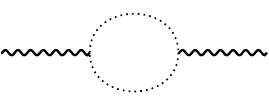}\label{vector_g}}

    \caption{Gauge boson self-energy diagrams.}
    \label{vectorself}
\end{figure}

In this section, we present the full one-loop thermal mass corrections for the bosonic degrees of freedom. Figure~\ref{scalarself} shows the possible self-energy diagrams for scalar particles \{\(h,G,A,H,H^\pm\)\}, and Fig.~\ref{vectorself} shows those for gauge bosons \{\(W^a,B\)\}, where subfigure~\ref{vector_g} represents the ghost contribution to the gauge boson self-energy.

In the Landau gauge, the gauge boson propagator reads
\begin{align}
D^{\mu\nu}(p) = \frac{-1}{p^2 - m^2_{T}(\phi)}T^{\mu\nu} + \frac{-1}{p^2 - m^2_{L}(\phi)}L^{\mu\nu},
\end{align}
where the first term on the right hand side corresponds to the transverse component and the second to the longitudinal component. The projection operators are given by
\begin{align}
T_{\mu\nu} &= g_\mu^{\ i}\left(\delta_{ij} - \frac{p_ip_j}{\vec{p}^2} \right)g^j_{\ \nu},\\
L_{\mu\nu} &=  \frac{p_\mu p_\nu}{p^2} - g_{\mu\nu} - T_{\mu\nu},
\end{align}
which satisfy the identities
\begin{align}
  T^{\mu\rho}T_{\rho\nu} &= -T^{\mu}_{\ \ \nu},\\
L^{\mu\rho}L_{\rho\nu} &= -L^{\mu}_{\ \, \nu},\\
T^{\mu\rho}L_{\rho\nu} &= 0,\\
T^\mu_{\ \ \mu} &= -(2-2\epsilon),\\
L^\mu_{\ \, \mu} &= -1.  
\end{align}
For a generic gauge boson self-energy polarization tensor \(\Pi_{\mu\nu}(p)\), we can decompose it into transverse and longitudinal components using these projection operators:
\begin{equation}
\Pi_{\mu\nu}(p) = \Pi_T(p)T_{\mu\nu} + \Pi_L(p) L_{\mu\nu} .
\end{equation}
Note that \(p^\mu = (i\omega_n, \ \vec{p})\) denotes the Minkowski momentum.
In the limit of vanishing external momentum, the vertices involved in the self-energy diagram~\ref{scalar_e} are proportional to \(p^\mu\). Since \(p^\mu T_{\mu\nu}=p^\mu L_{\mu\nu}=0\), the contributions of such diagrams vanish.

In what follows, we present the thermal mass corrections for each scalar particle and express them in terms of the functions \(I_0^{b/f}(m^2,T)\) and \(I_j^{b/f}(m^2,T)\) defined by Eqs.~\eqref{eq:Ibf0} and \eqref{eq:Ijfi}.
{\allowdisplaybreaks
\begin{itemize}[leftmargin=12pt, nosep]
    \item Momentum-independent self-energy of the Higgs:
    \begin{align}
    \Pi_h^h &= \frac{3}{2}\lambda_1I_1^b(m_h^2,T), \\
    \Pi_h^A &= \frac{1}{2}\lambda_-I_1^b(m_A^2,T) ,\\
    \Pi_h^H &= \frac{1}{2}\lambda_+I_1^b(m_H^2,T) ,\\
    \Pi_h^{H^\pm} &= \lambda_3I_1^b(m_{H^\pm}^2,T) ,\\
    \Pi_h^{G} &= \frac{3}{2}\lambda_1I_1^b(m_{G}^2,T) ,\\
    \Pi_h^{Z} &= -\frac{1}{2}\sint_K\frac{(gc_w+g^\prime s_w)^2}{2}g_{\mu\nu}D_Z^{\mu\nu}(k) \nonumber\\
       &= \frac{1}{4} \left[(gc_{w,L}+g^\prime s_{w,L})^2I_1^b(m_{Z,L}^2,T) + (gc_{w,T}+g^\prime s_{w,T})^2(2-2\epsilon)I_1^b(m_{Z,T}^2,T) \right],\\
    \Pi_h^W  &= \frac{1}{2}g^{ 2} \left[I_1^b(m_{W,L}^2,T) + (2-2\epsilon)I_1^b(m_{W,T}^2,T) \right] ,\\
    \Pi_h^{\gamma} &= \frac{1}{4} \left[(gs_{w,L}-g^\prime c_{w,L})^2I_1^b(m_{\gamma,L}^2,T) + (gs_{w,T}-g^\prime c_{w,T})^2(2-2\epsilon)I_1^b(m_{\gamma,T}^2,T) \right].
    \end{align}
    Here the sum-integral symbol \({\scriptstyle\Sigma}\!\!\!\!\int_K\) is defined in Eq.~\eqref{eq:sum-int}.
    \item Momentum-dependent self-energy of the Higgs:
    \begin{align}
    \Pi^{tt}_h(0) &= \frac{3y^2_t}{2}\sint_K\frac{\mathrm{tr}(\slashed{k}+m_t)^2}{(k^2-m_t^2)^2}\nonumber\\
                  &= 6y_t^2[-I^f_1(m_t^2,T) + 2m_t^2I^f_2(m^2_t,T)],\\
    \Pi_h^{hh}(0) &= - \frac{9}{2}\lambda_1^2\phi^2I_2^b(m_h^2,T),\\
    \Pi_h^{AA}(0) &= - \frac{1}{2}\lambda_-^2\phi^2I_2^b(m_A^2,T),\\
    \Pi_h^{HH}(0) &= - \frac{1}{2}\lambda_+^2\phi^2I_2^b(m_H^2,T),\\
    \Pi_h^{H^+ H^-}(0) &= - \lambda_3^2\phi^2I_2^b(m_{H^\pm}^2,T),\\
    \Pi_h^{GG}(0) &= - \frac{3}{2}\lambda_1^2\phi^2I_2^b(m_{G}^2,T),\\
    \Pi_h^{ZZ}(0) &= -\frac{1}{2}\sint_K\frac{(gc_w+g^\prime s_w)^4\phi^2}{4}g_{\mu\nu}D_Z^{\mu\rho}(k)g_{\rho\sigma}D_Z^{\nu\sigma}(k) \nonumber\\
       &= - \frac{1}{8}\phi^2 \left[(gc_{w,L}+g^\prime s_{w,L})^4I_2^b(m_{Z,L}^2,T) + (gc_{w,T}+g^\prime s_{w,T})^4(2-2\epsilon)I_2^b(m_{Z,T}^2,T) \right],\\
    \Pi_h^{WW}(0)  &=  - \frac{1}{4}g^{ 4}\phi^2 \left[I_2^b(m_{W,L}^2,T) + (2-2\epsilon)I_2^b(m_{W,T}^2,T) \right],\\
    \Pi_h^{\gamma\gamma}(0) &=- \frac{1}{8}\phi^2 \left[(gs_{w,L}-g^\prime c_{w,L})^4I_2^b(m_{\gamma,L}^2,T) + (gs_{w,T}-g^\prime c_{w,T})^4 (2-2\epsilon)I_2^b(m_{\gamma,T}^2,T) \right].
    \end{align}
    The explicit momentum dependence of the self-energy of the Higgs boson is obtained by substituting \(m_i^2\to m_i^2 - x(1-x)p^2\) into the above results and integrating over \(x\) from 0 to 1.

    \item Momentum-independent self-energy of the Goldstone:
    \begin{align}
    \Pi_G^{h} &= \frac{1}{2}\lambda_1 I_1^b(m_h^2,T) ,\\
   \Pi_G^G &= \frac{5}{2}\lambda_1I_1^b(m_G^2,T),\\
   \Pi_G^A &= \frac{1}{2}\lambda_+ I_1^b(m_A^2,T),\\
   \Pi_G^H &= \frac{1}{2}\lambda_- I_1^b(m_H^2,T),\\
   \Pi_G^{H^\pm} &= \lambda_3 I_1^b(m_{H^\pm}^2,T),\\
   \Pi_G^Z &= \frac{1}{4}\left[(gc_{w,L}+g^\prime s_{w,L})^2I_1^b(m_{Z,L}^2,T) + (gc_{w,T}+g^\prime s_{w,T})^2(2-2\epsilon)I_1^b(m_{Z,T}^2,T)\right],\\
   \Pi_G^{W} &= \frac{1}{2}g^2 \left[I_1^b(m_{W,L}^2,T) + (2-2\epsilon)I_1^b(m_{W,T}^2,T)\right],\\
   \Pi_G^\gamma &= \frac{1}{4}\left[(gs_{w,L}-g^\prime c_{w,L})^2I_1^b(m_{\gamma,L}^2,T) + (gs_{w,T}-g^\prime c_{w,T})^2(2-2\epsilon)I_1^b(m_{\gamma,T}^2,T)\right].
    \end{align}
    \item Momentum-dependent self-energy of the Goldstone:
    \begin{align}
         \Pi_G^{tt}(0) &=  -\frac{3y^2_t}{2}\sint_K\frac{\mathrm{tr}[\gamma^5(\slashed{k}+m_t)\gamma^5(\slashed{k}+m_t)]}{(k^2-m_t^2)^2}\nonumber\\
    &= -6y_t^2I^f_1(m_t^2,T),\\
    \Pi_G^{hG}(0) &= -\lambda_1^2\phi^2 I_{\rm mix}(m_h^2,m_G^2,T),\\
    \Pi_G^{AH}(0 ) &= -\lambda_5^2\phi^2 I_{\rm mix} (m_A^2,m_H^2,T) ,
    \end{align}
    where we define
    \begin{align}
 I_{\rm mix}(m_1^2,m_2^2,T) 
     & = \sint_K\frac{1}{(\omega_{n_b}^2 + k^2 + m_1^2)(\omega_{n_b}^2 + k^2 + m_2^2)}\nonumber\\
     &= \frac{1}{m_2^2 - m_1^2}\sint_K\left(\frac{1}{\omega_{n_b}^2 + k^2 + m_1^2}-\frac{1}{\omega_{n_b}^2 + k^2 + m_2^2}\right)\nonumber\\
     &= \frac{1}{m_2^2 - m_1^2}\left[I_1^b(m_1^2,T) -I_1^b(m_2^2,T)\right].
    \end{align}
    \item Momentum-independent self-energy of the CP-odd scalar:
    \begin{align}
    \Pi_A^{h} &= \frac{1}{2}\lambda_-I_1^b(m_h^2,T) ,\\
     \Pi_A^G &=  \left(\frac{1}{2}\lambda_+ +\lambda_3 \right)I_1^b(m_G^2,T),\\
     \Pi_A^A &= \frac{3}{2}\lambda_2 I_1^b(m_A^2,T),\\
     \Pi_A^H &= \frac{1}{2}\lambda_2 I_1^b(m_H^2,T),\\
     \Pi_A^{H^\pm} &= \lambda_2 I_1^b(m_{H^\pm}^2,T),\\
     \Pi_A^Z &= \frac{1}{4} \left[(gc_{w,L}+g^\prime s_{w,L})^2I_1^b(m_{Z,L}^2,T) + (gc_{w,T}+g^\prime s_{w,T})^2(2-2\epsilon)I_1^b(m_{Z,T}^2,T)\right],\\
     \Pi_A^W &= \frac{1}{2}g^2 \left[I_1^b(m_{W,L}^2,T) + (2-2\epsilon)I_1^b(m_{W,T}^2,T)\right],\\
     \Pi_A^\gamma &= \frac{1}{4} \left[(gs_{w,L}-g^\prime c_{w,L})^2I_1^b(m_{\gamma,L}^2,T) + (gs_{w,T}-g^\prime c_{w,T})^2(2-2\epsilon)I_1^b(m_{\gamma,T}^2,T)\right].
\end{align}

    \item Momentum-dependent self-energy of the CP-odd scalar:
    \begin{align}
         \Pi_A^{Ah}(0) &= -\lambda_-^2\phi^2 I_{\rm mix}(m_A^2,m_h^2,T),\\
     \Pi_A^{HG^0}(0) &= -\lambda_5^2\phi^2 I_{\rm mix}(m_H^2,m_G^2,T),\\
     \Pi_A^{H^\pm G^\mp}(0) &= -\frac{1}{2}(\lambda_4-\lambda_5)^2\phi^2 I_{\rm mix} (m_{H^\pm}^2,m_G^2,T).
    \end{align}

    \item Momentum-independent self-energy of the CP-even scalar:
    \begin{align}
        \Pi_H^h &= \frac{1}{2}\lambda_+ I_1^b(m_h^2,T),\\
    \Pi_H^G &= \left(\frac{1}{2}\lambda_- +\lambda_3\right) I_1^b(m_G^2,T),\\
    \Pi_H^A &= \frac{1}{2}\lambda_2 I_1^b(m_A^2,T),\\
    \Pi_H^H &= \frac{3}{2}\lambda_2 I_1^b(m_H^2,T),\\
    \Pi_H^{H^\pm} &= \lambda_2 I_1^b(m_{H^\pm}^2,T),\\
    \Pi_H^Z &= \frac{1}{4} \left[(gc_{w,L}+g^\prime s_{w,L})^2I_1^b(m_{Z,L}^2,T) + (gc_{w,T}+g^\prime s_{w,T})^2(2-2\epsilon)I_1^b(m_{Z,T}^2,T)\right],\\
     \Pi_H^W &= \frac{1}{2}g^2 \left[I_1^b(m_{W,L}^2,T) + (2-2\epsilon)I_1^b(m_{W,T}^2,T)\right],\\
     \Pi_H^\gamma &= \frac{1}{4} \left[(gs_{w,L}-g^\prime c_{w,L})^2I_1^b(m_{\gamma,L}^2,T) + (gs_{w,T}-g^\prime c_{w,T})^2(2-2\epsilon)I_1^b(m_{\gamma,T}^2,T)\right].
    \end{align}

    \item Momentum-dependent self-energy of the CP-even scalar:
    \begin{align}
        \Pi_H^{Hh}(0) &= -\lambda_+^2\phi^2 I_{\rm mix}(m_H^2,m_h^2,T),\\
    \Pi_H^{AG}(0) &= -\lambda_5^2\phi^2 I_{\rm mix}(m_A^2,m_G^2,T),\\
    \Pi_H^{H^\pm G^\mp}(0) &= -\frac{1}{2}(\lambda_4+\lambda_5)^2\phi^2 I_{\rm mix} (m_{H^\pm}^2,m_G^2,T).
    \end{align}

    \item Momentum-independent self-energy of the charged scalar:
    \begin{align}
     \Pi_{H^\pm}^h &= \frac{1}{2}\lambda_3 I_1^b(m_h^2,T),\\
      \Pi_{H^\pm}^G &= \left(\frac{3}{2}\lambda_3+\lambda_4 \right)I_1^b(m_G^2,T),\\
      \Pi_{H^\pm}^A &= \frac{1}{2}\lambda_2 I_1^b(m_A^2,T),\\
      \Pi_{H^\pm}^H &= \frac{1}{2}\lambda_2 I_1^b(m_H^2,T),\\
      \Pi_{H^\pm}^{H^\pm} &= 2\lambda_2 I_1^b(m_{H^\pm}^2,T),\\
      \Pi_{H^\pm}^{Z} &= \frac{1}{4}\left[(gc_{w,L}-g^\prime s_{w,L})^2I_1^b(m_{Z,L}^2,T)+(gc_{w,T}-g^\prime s_{w,T})^2(2-2\epsilon)I_1^b(m_{Z,T}^2,T)\right],\\
       \Pi_{H^\pm}^{W} &=\frac{1}{2}g^2 \left[I_1^b(m_{W,L}^2,T)+(2-2\epsilon)I_1^b(m_{W,T}^2,T)\right],\\
        \Pi_{H^\pm}^{\gamma} &= \frac{1}{4}\left[(gs_{w,L}+g^\prime c_{w,L})^2I_1^b(m_{\gamma,L}^2,T)+(gs_{w,T}+g^\prime c_{w,T})^2(2-2\epsilon)I_1^b(m_{\gamma,T}^2,T)\right].
\end{align}

    \item Momentum-dependent self-energy of the charged scalar:
    \begin{align}
        \Pi_{H^\pm}^{H^\mp h}(0) &= -\lambda_3^2\phi^2 I_{\rm mix}(m_{H^\pm}^2,m_h^2,T),
      \\
      \Pi_{H^\pm}^{AG^\mp}(0) &= -\frac{1}{4}(\lambda_4-\lambda_5)^2\phi^2 I_{\rm mix}(m_A^2,m_{G}^2,T),\\
      \Pi_{H^\pm}^{HG^\mp}(0) &= -\frac{1}{4}(\lambda_4+\lambda_5)^2\phi^2 I_{\rm mix}(m_H^2,m_{G}^2,T),\\
    \end{align}

    \item Self-energy of the gauge bosons: 
    
    Like \(\mathcal{M}^2\), \(\Pi_{T/L}(p)\) forms a \(4\times 4\) matrix. With the projection operator identities, we evaluate \(\Pi_{T/L}(p)\) via 
\begin{align}
    \Pi_T(p) = \frac{1}{2-2\epsilon}T_{\mu\nu}(p) \Pi^{\mu\nu}(p),\quad \Pi_L(p) = L_{\mu\nu}(p) \Pi^{\mu\nu}(p).
\end{align}
It is useful to define
\begin{align}
     D_T I_0^{b/f}(m^2,T) &=-\frac{1}{2} \sint_K \ln(K^2 + m^2)-  \sint_K \frac{\omega_n^2}{K^2 + m^2},\\
     D_T I_j^{b/f}(m^2,T) &= \frac{1}{2} \sint_K \frac{1}{(K^2 + m^2)^j} - j  \sint_K \frac{\omega_n^2}{(K^2 + m^2)^{j+1}},
\end{align}
where the operator \(D_T\equiv T^2\mathrm{d}/\mathrm{d}T^2\). Furthermore, a useful substitution for evaluating loop-momentum integrals is \(k^i k^j \to -g^{ij}k^2/(3-2\epsilon)\).

The momentum-independent contributions to the (11) and (22) components of \(\Pi_T\) and \(\Pi_L\) are given by
    \begin{align}
         \Pi_{T,11}^{h} &= \Pi_{T,22}^{h} =\Pi_{L,11}^h=\Pi_{L,22}^h=  \frac{g^2}{4}I^b_1(m_h^2,T),\\
     \Pi_{T,11}^G &= \Pi_{T,22}^G =  \Pi_{L,11}^G = \Pi_{L,22}^G = \frac{3g^2}{4}I^b_1(m_G^2,T),\\
      \Pi_{T,11}^A &= \Pi_{T,22}^A =\Pi_{L,11}^A =\Pi_{L,22}^A =  \frac{g^2}{4}I^b_1(m_A^2,T),\\
       \Pi_{T,11}^H &= \Pi_{T,22}^H=\Pi_{L,11}^H =  \Pi_{L,22}^H = \frac{g^2}{4}I^b_1(m_H^2,T),\\
        \Pi_{T,11}^{H^\pm} &=\Pi_{T,22}^{H^\pm} = \Pi_{L,11}^{H^\pm} = \Pi_{L,22}^{H^\pm} = \frac{g^2}{2}I^b_1(m_{H^\pm}^2,T),\\
        \Pi^Z_{T,11} &= \Pi^Z_{T,22}=g^2 \Pi_T^V(m^2_{Z,L}, m^2_{Z,T}, T, c_{w,L}, c_{w,T}),\\
         \Pi^Z_{L,11} &=\Pi^Z_{L,22}=g^2 \Pi_L^V(m^2_{Z,L}, m^2_{Z,T}, T, c_{w,L}, c_{w,T}),\\
         \Pi^W_{T,11} &=\Pi^W_{T,22}= g^2 \Pi_T^V(m^2_{W,L}, m^2_{W,T}, T, 1, 1),\\
         \Pi^W_{L,11} &=\Pi^W_{L,22} =g^2 \Pi_L^V(m^2_{W,L}, m^2_{W,T}, T, 1,1),\\
         \Pi^\gamma_{T,11} &= \Pi^\gamma_{T,22}=g^2 \Pi_T^V(m^2_{\gamma,L}, m^2_{\gamma,T}, T, s_{w,L}, s_{w,T}),\\
         \Pi^\gamma_{L,11} &=\Pi^\gamma_{L,22}=g^2 \Pi_L^V(m^2_{\gamma,L}, m^2_{\gamma,T}, T, s_{w,L}, s_{w,T}),
    \end{align}
 where the polarization function \(\Pi^V_{L/T}\) from vector boson contribution is defined as
 \begin{align}
     &\Pi_T^V(m^2_{L}, m^2_{T}, T, c_{w,L}, c_{w,T}) \equiv -\frac{c_w^2}{2(2-2\epsilon)} T^{\mu\nu}\sint_k(g_{\mu\rho}g_{\nu\sigma}+g_{\mu\sigma}g_{\nu\rho} - 2g_{\mu\nu}g_{\rho\sigma}) D_Z^{\rho\sigma}(k)  \nonumber\\
     &= \frac{4}{3}c^2_{w,T}\left(1-\frac{4}{3}\epsilon\right) I_{1}^b(m^2_{T},T) +  \frac{c^2_{w,L}}{m^2_L} \left[\left(\frac{1}{6}+\frac{1}{9}\epsilon\right)I_0^b(m^2_L, T) - \frac{1}{3}D_TI_0^b(m^2_L, T)\right.\nonumber\\
     &\left.\ \ \  - \frac{1}{6}I_0^b(0, T) + \frac{1}{3} D_TI_0^b(0, T) \right] +c^2_{w,L} I_1^b(m^2_L, T),\\
     &\Pi_{L}^{V}(m^2_{L}, m^2_{T}, T, c_{w,L}, c_{w,T}) \equiv -\frac{c_w^2}{2} L^{\mu\nu}\sint_k(g_{\mu\rho}g_{\nu\sigma}+g_{\mu\sigma}g_{\nu\rho} - 2g_{\mu\nu}g_{\rho\sigma}) D_Z^{\rho\sigma}(k)\nonumber\\
     &= c^2_{w,T}(2-2\epsilon)I_1^b(m^2_T,T) - \frac{c^2_{w,L}}{2m_L^2 } \left[ I_0^b(m_L^2,T)  - 2D_TI_0^b(m_L^2,T) - I_0^b(0,T) + 2D_TI_0^b(0,T) \right].
 \end{align}
The momentum-dependent contributions to the (11) and (22) components of \(\Pi_T\) and \(\Pi_L\) are given by
\begin{align}
    \Pi_{T,11}^{ff}(0) &= \Pi_{T,22}^{ff}(0) = \frac{3}{2(2-2\epsilon)}g^2\sint_K\frac{T_{\mu\nu}\mathrm{tr}[\gamma^\mu P_L (\slashed{k}+m_t)\gamma^\nu P_L\slashed{k}]}{k^2(k^2-m_t^2)}\nonumber\\
    &\ \ \  + \frac{9}{2(2-2\epsilon)}g^2\sint_K\frac{T_{\mu\nu}\mathrm{tr}[\gamma^\mu P_L \slashed{k}\gamma^\nu P_L\slashed{k}]}{k^4}\nonumber\\
    &= \frac{g^2}{m_t^2}\left[\left(1+\frac{1}{6}\epsilon\right) I_0^f(m^2_t,T) - I_0^f(0,T) - 2 D_TI_0^f(m^2_t,T) + 2 D_TI_0^f(0,T)\right]  \nonumber\\
    &\ \ \ -g^2\left(1-\frac{11}{6}\epsilon\right) I_1^f(m^2_t,T),\\
     \Pi_{L,11}^{ff}(0) &= \Pi_{L,22}^{ff}(0) = \frac{3}{2(2-2\epsilon)}g^2\sint_K\frac{L_{\mu\nu}\mathrm{tr}[\gamma^\mu P_L (\slashed{k}+m_t)\gamma^\nu P_L\slashed{k}]}{k^2(k^2-m_t^2)}\nonumber\\
    &\ \ \  + \frac{9}{2(2-2\epsilon)}g^2\sint_K\frac{L_{\mu\nu}\mathrm{tr}[\gamma^\mu P_L \slashed{k}\gamma^\nu P_L\slashed{k}]}{k^4}\nonumber\\
    &= \frac{3g^2}{m_t^2}\left[-\left(1-\frac{1}{2}\epsilon\right) I_0^f(m^2_t,T) +I_0^f(0,T) + 2 D_TI_0^f(m^2_t,T) - 2 D_TI_0^f(0,T) \right]  \nonumber\\
    &\ \ \ - 3g^2\left[\left(1-\frac{1}{2}\epsilon\right)I_1^f(m_t^2,T) + 6I_1^f(0,T)\right],\\
    \Pi_{T,11}^{hG}(0) &= \Pi_{T,22}^{hG}(0)=-g^2 \Pi_T^{SS}(m_h^2,m_G^2,T),\\
    \Pi_{L,11}^{hG}(0) &= \Pi_{L,22}^{hG}(0)=-g^2 \Pi_L^{SS}(m_h^2,m_G^2,T),\\
    \Pi_{T,11}^{HH^\pm}(0) &=\Pi_{T,22}^{HH^\pm}(0)= -g^2 \Pi_T^{SS} (m_H^2,m_{H^\pm}^2,T),\\
    \Pi_{L,11}^{HH^\pm}(0) &=\Pi_{L,22}^{HH^\pm}(0)= -g^2 \Pi_L^{SS} (m_H^2,m_{H^\pm}^2,T),\\
    \Pi_{T,11}^{AH^\pm}(0) &=\Pi_{T,22}^{AH^\pm}(0)= -g^2 \Pi_T^{SS} (m_A^2,m_{H^\pm}^2,T),\\
    \Pi_{L,11}^{AH^\pm}(0) &= \Pi_{L,22}^{AH^\pm}(0)= -g^2 \Pi_L^{SS} (m_A^2,m_{H^\pm}^2,T),\\
    \Pi_{T,11}^{GG}(0) &= \Pi_{T,22}^{GG}(0)= -\frac{g^2}{2-2\epsilon}\sint_k \frac{T_{\mu\nu}k^\mu k^\nu}{(k^2-m_G^2)(k^2-m_G^2)}\nonumber\\
    &=\frac{g^2}{3}\left[-\frac{1}{2}\left(1+\frac{2}{3}\epsilon\right)I_1^b(m_G^2,T) + m_G^2 \left(1+\frac{2}{3}\epsilon\right)I_2^b(m_G^2,T)  - D_T I_1^b(m_G^2,T)  \right],\\
    \Pi_{L,11}^{GG}(0) &= \Pi_{L,22}^{GG}(0)= -g^2\sint_k \frac{L_{\mu\nu}k^\mu k^\nu}{(k^2-m_G^2)(k^2-m_G^2)}\nonumber\\
    &= -\frac{g^2}{2}\left[I_1^b(m_G^2,T) - 2D_T I_1^b(m_G^2,T) \right]  ,\\
    \Pi_{T,11}^{hW}(0) &= \Pi_{T,22}^{hW}(0)=-\frac{g^4\phi^2}{4}\Pi_T^{SV}(m_h^2,m_{W,L}^2,m_{W,T}^2,T, 1,1),\\
    \Pi_{L,11}^{hW}(0) &= \Pi_{L,22}^{hW}(0)=-\frac{g^4\phi^2}{4}\Pi_L^{SV}(m_h^2,m_{W,L}^2,T),\\
    \Pi_{T,11}^{GZ}(0) &= \Pi_{T,22}^{GZ}(0)=-\frac{g^2 g^{\prime 2}\phi^2}{4}\Pi_T^{SV}(m_G^2,m_{Z,L}^2,m_{Z,T}^2,T,s_{w,L},s_{w,T}),\\
    \Pi_{L,11}^{GZ}(0) &= \Pi_{L,22}^{GZ}(0)=-\frac{g^2 g^{\prime 2}s_{w,L}^2\phi^2}{4}\Pi_L^{SV}(m_G^2,m_{Z,L}^2,T),\\
    \Pi_{T,11}^{G\gamma}(0) &= \Pi_{T,22}^{G\gamma}(0)=-\frac{g^2 g^{\prime 2}\phi^2}{4}\Pi_T^{SV}(m_G^2,m_{\gamma,L}^2,m_{\gamma,T}^2,T,c_{w,L},c_{w,T}),\\
    \Pi_{L,11}^{G\gamma}(0) &= \Pi_{L,22}^{G\gamma}(0)=-\frac{g^2 g^{\prime 2}c_{w,L}^2\phi^2}{4}\Pi_L^{SV}(m_G^2,m_{\gamma,L}^2,T),\\
    \Pi_{T,11}^{W\gamma}(0) &= \Pi_{T,22}^{W\gamma}(0)\nonumber\\
    &=-\frac{g^2s_w^2}{2-2\epsilon} T^{\mu\nu}\sint_k(g_{\mu\alpha}k_\rho- 2g_{\alpha\rho}k_\mu + g_{\rho\mu}k_\alpha)D_\gamma^{\rho\sigma}(g_{\beta\nu}k_\sigma+g_{\nu\sigma}k_\beta -2g_{\sigma\beta}k_\nu)D_W^{\alpha\beta} \nonumber\\
    &= 4g^2\left[ \frac{1}{3}s^2_{w,L} \Pi_T^{VV}(m^2_{W,L}, m^2_{\gamma,L},T,2/3) +\frac{2}{3}s^2_{w,T} \Pi_T^{VV}(m^2_{W,T}, m^2_{\gamma,T},T,-1/3) \right],\\
    \Pi_{L,11}^{W\gamma}(0) &= \Pi_{L,22}^{W\gamma}(0)\nonumber\\
    &=2g^2\left[s^2_{w,L} \Pi_L^{VV}(m^2_{W,L}, m^2_{\gamma,L},T,0) + 2s^2_{w,T} \Pi_L^{VV}(m^2_{W,T}, m^2_{\gamma,T},T,-1) \right]        ,\\
    \Pi_{T,11}^{WZ}(0) &= \Pi_{T,22}^{WZ}(0)\nonumber\\
    &=4g^2\left[\frac{2}{3}c^2_{w,T} \Pi_T^{VV}(m^2_{W,T}, m^2_{Z,T},T,-1/3) + \frac{1}{3}c^2_{w,L} \Pi_T^{VV}(m^2_{W,L}, m^2_{Z,L},T,2/3) \right],\\
    \Pi_{L,11}^{WZ}(0) &= \Pi_{L,22}^{WZ}(0)\nonumber\\
    &=2g^2\left[c^2_{w,L} \Pi_L^{VV}(m^2_{W,L}, m^2_{Z,L},T,0) + 2c^2_{w,T} \Pi_L^{VV}(m^2_{W,T}, m^2_{Z,T},T,-1) \right]        ,\\
    \Pi_{T,11}^{cc}(0) &= \Pi_{T,22}^{c}(0) = g^2I_1^b(0,T),\\
    \Pi_{L,11}^{cc}(0) &= \Pi_{L,22}^{c}(0) = -g^2I_1^b(0,T),
\end{align}
where we define
\begin{align}
    &\Pi^{SS}_T(m^2_1,m^2_2,T) \equiv  \frac{1}{2-2\epsilon}\sint_k \frac{T_{\mu\nu}k^\mu k^\nu}{(k^2-m_1^2)(k^2-m_2^2)} \nonumber\\
    &= \frac{1}{3}\left(1+\frac{2}{3}\epsilon\right)\left[I_1^b(m_2^2, T) - m_1^2  I_{\rm mix}(m_1^2, m^2_2, T) \right]\nonumber\\
    &\ \ \  + \frac{1}{3(m^2_2 - m^2_1)}\left(1+\frac{2}{3}\epsilon\right)\left[\frac{1}{2} I_0^b(m^2_2, T) - D_TI_0^b(m_2^2, T) - \frac{1}{2} I_0^b(m_1^2, T) + D_TI_0^b(m_1^2, T)\right],\\
    &\Pi_{L}^{SS}(m^2_1,m^2_2,T)\equiv \sint_k \frac{L_{\mu\nu}k^\mu k^\nu}{(k^2-m_G^2)(k^2-m_h^2)}\nonumber\\
      &= \frac{1}{m_2^2 - m_1^2}\left[\frac{1}{2}I_0^b(m_1^2,T)  - D_TI_0^b(m_1^2,T) - \frac{1}{2}I_0^b(m_2^2,T) + D_TI_0^b(m_2^2,T)\right],\\
      &\Pi_T^{SV}(m^2_s,m^2_L,m^2_T,T,s_L,s_T) \equiv \frac{s^2}{(2-2\epsilon)}\sint_k  \frac{T_{\mu\nu}g^{\mu \rho} g^{\sigma \nu}}{(k^2 - m_s^2)(k^2- m^2)} \left( \frac{k_\rho k_\sigma}{k^2} - g_{\rho\sigma}\right)\nonumber\\
      &= \frac{2}{3}\left(1-\frac{1}{3}\epsilon\right)s_T^2I_{\rm mix}(m^2_T, m_s^2, T) \nonumber\\
      &\ \ \ - \frac{s_L^2}{3m_s^2(m_s^2 - m_L^2)}\left[-\frac{1}{2}\left(1+\frac{2}{3}\epsilon\right)I_0^b(m_s^2,T) + D_T I_0^b(m_s^2,T) - \frac{3}{2}I_0^b(0,T) \right]\nonumber\\
   &\ \ \ + \frac{s_L^2}{3m_L^2(m_s^2 - m_L^2)}\left[-\frac{1}{2}\left(1+\frac{2}{3}\epsilon\right)I_0^b(m_L^2,T) + D_T I_0^b(m_L^2,T) - \frac{3}{2}I_0^b(0,T) \right] ,\\
  & \Pi_L^{SV}(m_s^2,m_{L}^2,T) \equiv \sint_k  \frac{L^{\rho\sigma}}{(k^2 - m_s^2)(k^2- m_L^2)} \left( \frac{k_\rho k_\sigma}{k^2} - g_{\rho\sigma}\right) \nonumber\\
  &= I_{\rm mix}(m_L^2,m_s^2,T)  + \frac{1}{m_s^2(m_s^2 - m_L^2)}\left[-\frac{1}{2}I_0^b(m_s^2,T) + D_T I_0^b(m_s^2,T) - \frac{3}{2}I_0^b(0,T) \right]\nonumber\\
   &\ \ \ - \frac{1}{m_L^2(m_s^2 - m_L^2)}\left[-\frac{1}{2}I_0^b(m_L^2,T) + D_T I_0^b(m_L^2,T) - \frac{3}{2}I_0^b(0,T) \right],\\
  & \Pi_T^{VV}(m^2_1,m^2_2,T,\kappa)\equiv \left(1+\kappa \epsilon\right)\left[-I_1^b(m_1^2,T)  + m_2^2I_{\rm mix}(m_1^2,m_2^2,T) \right] \nonumber\\
   & \qquad\quad - \frac{ 1+\kappa \epsilon}{m_1^2 - m_2^2}\left[\frac{1}{2}I_0^b(m_1^2,T)  - D_TI_0^b(m_1^2,T) - \frac{1}{2}I_0^b(m_2^2,T) + D_TI_0^b(m_2^2,T)\right],\\
    & \Pi_L^{VV}(m^2_1,m^2_2,T,\kappa)\equiv \frac{1+\kappa\epsilon}{m_1^2 - m_2^2}\left[I_0^b(m_1^2,T)   - I_0^b(m_2^2,T)  - 2D_TI_0^b(m_1^2,T) + 2D_TI_0^b(m_2^2,T)\right].
\end{align}
The momentum-independent contributions to the (33) component of \(\Pi_T\) and \(\Pi_L\) are given by
\begin{align}
     \Pi_{T,33}^{h} &=\Pi_{L,33}^h=  \frac{g^2}{4}I^b_1(m_h^2,T),\\
     \Pi_{T,33}^G &= \Pi_{L,33}^G =  \frac{3g^2}{4}I^b_1(m_G^2,T),\\
      \Pi_{T,33}^A &= \Pi_{L,33}^A =  \frac{g^2}{4}I^b_1(m_A^2,T),\\
       \Pi_{T,33}^H &= \Pi_{L,33}^H =  \frac{g^2}{4}I^b_1(m_H^2,T),\\
        \Pi_{T,33}^{H^\pm} &= \Pi_{L,33}^{H^\pm} =  \frac{g^2}{2}I^b_1(m_{H^\pm}^2,T),\\
         \Pi_{T,33}^W &= 2\Pi^W_{T,11},\\
         \Pi_{L,33}^W &= 2\Pi^W_{L,11}.
\end{align}
The momentum-dependent contributions to the (33) component of \(\Pi_T\) and \(\Pi_L\) are given by
\begin{align}
    \Pi_{T,33}^{ff}(0) &= \frac{3}{4(2-2\epsilon)}g^2T_{\mu\nu}\sint_k\left(\frac{\mathrm{tr}[\gamma^\mu  P_L(\slashed{k}+m_t)\gamma^\nu P_L (\slashed{k}+m_t)]}{(k^2-m_t^2)^2}  + 7\sint_k\frac{\mathrm{tr}[\gamma^\mu  P_L\slashed{k}\gamma^\nu P_L \slashed{k}]}{k^4}\right)\nonumber\\
    &= \frac{1}{2}g^2\left[-\left(2-\frac{5}{3}\epsilon\right)I_1^f(m_t^2,T)  +  2D_TI_1^f(m_t^2,T)    + \left(1-\frac{11}{6}\epsilon\right)  m_t^2 I_2^f(m_t^2,T) \right],\\
     \Pi_{L,33}^{ff}(0) &=\frac{3}{4}g^2L_{\mu\nu}\sint_k\left(\frac{\mathrm{tr}[\gamma^\mu  P_L(\slashed{k}+m_t)\gamma^\nu P_L (\slashed{k}+m_t)]}{(k^2-m_t^2)^2}  + 7\sint_k\frac{\mathrm{tr}[\gamma^\mu  P_L\slashed{k}\gamma^\nu P_L \slashed{k}]}{k^4}\right)\nonumber\\
     &=\frac{3}{4}g^2\left[-4D_TI_1^f(m_t^2,T) -28D_TI_1^f(0,T)  + (2-\epsilon)m_t^2 I_2^f(m_t^2,T)  \right],\\
     \Pi_{T,33}^{hG}(0) &= \Pi_{T,11}^{hG}(0),\\
    \Pi_{L,33}^{hG}(0) &= \Pi_{L,11}^{hG}(0),\\
    \Pi_{T,33}^{GG}(0) &= \Pi_{T,11}^{GG}(0),\\
     \Pi_{L,33}^{GG}(0) &= \Pi_{L,11}^{GG}(0),\\
     \Pi_{T,33}^{AH}(0)  &= -g^2\Pi_{T}^{SS}(m^2_A,m^2_H,T),\\
     \Pi_{L,33}^{AH}(0)  &= -g^2\Pi_{L}^{SS}(m^2_A,m^2_H,T),\\
     \Pi_{T,33}^{H^+H^-}(0) &= \frac{g^2}{3}\left[-\frac{1}{2}\left(1+\frac{2}{3}\epsilon\right)I_1^b(m_{H^\pm}^2,T) + m_{H^\pm}^2 \left(1+\frac{2}{3}\epsilon\right)I_2^b(m_{H^\pm}^2,T)  - D_T I_1^b(m_{H^\pm}^2,T)  \right],\\
    \Pi_{L,33}^{H^+H^-}(0) &= -\frac{g^2}{2}\left[I_1^b(m_{H^\pm}^2,T) - 2D_T I_1^b(m_{H^\pm}^2,T) \right]  ,\\
    \Pi_{T,33}^{WW}(0) &= -4g^2\left[\frac{2}{3}\left(1-\frac{1}{3}\epsilon\right)\left(\frac{1}{2}I_1^b(m_{W,T}^2,T)  - m_{W,T}^2I_{2}^b(m_{W,T}^2,T)    + D_TI_1^b(m_{W,T}^2,T)\right) \right.\nonumber\\
    &\ \ \ \left. +\frac{1}{3}\left(1+\frac{2}{3}\epsilon\right)\left(\frac{1}{2}I_1^b(m_{W,L}^2,T)  - m_{W,L}^2I_{2}^b(m_{W,L}^2,T)    + D_TI_1^b(m_{W,L}^2,T)\right) \right],\\
    \Pi_{L,33}^{WW} (0)   &=-4g^2\left[\frac{1}{2}I_1^b(m_{W,L}^2,T)  - D_TI_1^b(m_{W,L}^2,T)+(1-\epsilon)I_1^b(m_{W,T}^2,T) - 2D_TI_1^b(m_{W,T}^2,T)\right],\\
    \Pi_{T,33}^{hZ}(0) &= -\frac{g^2\phi^2}{4}\Pi_T^{SV}(m_h^2,m^2_{Z,L},m^2_{Z,T},T,g c_{w,L} +  g^\prime s_{w,L}, g c_{w,T} +  g^\prime s_{w,T}) ,\\
    \Pi_{L,33}^{hZ}(0) &= -\frac{g^2(gc_{w,L} + g^\prime s_{w,L})^2\phi^2}{4}\Pi_L^{SV}(m_h^2,m^2_{Z,L},T),\\
    \Pi_{T,33}^{h\gamma}(0) &= -\frac{g^2\phi^2}{4}\Pi_T^{SV}(m_h^2,m^2_{\gamma,L},m^2_{\gamma,T},T,g s_{w,L} -  g^\prime c_{w,L}, g s_{w,T} -  g^\prime c_{w,T}) ,\\
    \Pi_{L,33}^{h\gamma}(0) &= -\frac{g^2(gs_{w,L} - g^\prime c_{w,L})^2\phi^2}{4}\Pi_L^{SV}(m_h^2,m^2_{\gamma,L},T),\\
   \Pi_{T,33}^{cc} (0) &= \Pi_{T,11}^{cc} (0) ,\\
   \Pi_{L,33}^{cc} (0) &= \Pi_{L,11}^{cc} (0) .
\end{align}
The momentum-independent contributions to the (44) component of \(\Pi_T\) and \(\Pi_L\) are given by
\begin{align}
 \Pi_{T,44}^{h}&=\Pi_{L,44}^h=  \frac{g^{\prime 2}}{4}I^b_1(m_h^2,T),\\
     \Pi_{T,44}^G&= \Pi_{L,44}^G =  \frac{3g^{\prime 2}}{4}I^b_1(m_G^2,T),\\
      \Pi_{T,44}^A&= \Pi_{L,44}^A =  \frac{g^{\prime 2}}{4}I^b_1(m_A^2,T),\\
       \Pi_{T,44}^H&= \Pi_{L,44}^H =  \frac{g^{\prime 2}}{4}I^b_1(m_H^2,T),\\
        \Pi_{T,44}^{H^\pm}&=\Pi_{L,44}^{H^\pm} =  \frac{g^{\prime 2}}{2}I^b_1(m_{H^\pm}^2,T).
\end{align}
The momentum-dependent contributions to the (44) component of \(\Pi_T\) and \(\Pi_L\) are given by
\begin{align}
    \Pi_{T,44}^{ff}(0) &= \frac{1}{12(2-2\epsilon)}g^{\prime 2}\sint_k \frac{T_{\mu\nu}\mathrm{tr}[\gamma^\mu (P_L + 4P_R)(\slashed{k}+m_t)\gamma^\nu (P_L+4P_R)(\slashed{k}+m_t)]}{(k^2-m_t^2)^2}\nonumber\\
    &\ \ \ + \frac{1}{6(2-2\epsilon)}g^{\prime 2}\sint_k \frac{T_{\mu\nu}\mathrm{tr}[\gamma^\mu (P_L + 4P_R)\slashed{k}\gamma^\nu (P_L+4P_R)\slashed{k}]}{k^4}\nonumber\\
    &\ \ \ + \frac{1}{4(2-2\epsilon)}g^{\prime 2}\sint_k \frac{T_{\mu\nu}\mathrm{tr}[\gamma^\mu (P_L - 2P_R)\slashed{k}\gamma^\nu (P_L-2P_R)\slashed{k}]}{k^4}\nonumber\\
    &\ \ \ + \frac{3}{4(2-2\epsilon)}g^{\prime 2}\sint_k \frac{T_{\mu\nu}\mathrm{tr}[\gamma^\mu (P_L + 2P_R)\slashed{k}\gamma^\nu (P_L+2P_R)\slashed{k}]}{k^4}\nonumber\\
     &\ \ \ + \frac{3}{4(2-2\epsilon)}g^{\prime 2}\sint_k \frac{T_{\mu\nu}\mathrm{tr}[\gamma^\mu P_L \slashed{k}\gamma^\nu P_L\slashed{k}]}{k^4}\nonumber\\
   &=  \frac{g^{\prime 2}}{12}\left\{ -\frac{34}{3}\left(1+\frac{1}{6}\epsilon\right) \left[I_1^f(m_t^2,T) - 2D_TI_1^f(m_t^2,T)\right] - 4(4-2\epsilon)m_t^2I_2^f(m_t^2,T)\right.  \nonumber\\
 &\ \ \ \left.+ \frac{34}{3}\left(1-\frac{11}{6}\epsilon\right) \left[-I_1^f(m_t^2,T)+ m_t^2I_2^f(m_t^2,T)  \right]    \right\},\\
\Pi_{L,44}^{ff}(0) &= \frac{g^{\prime 2}}{12}\left[ -34D_T I_1^f(m_t^2,T) + 9\left(2-\epsilon\right)m_t^2I_2^f(m_t^2,T)  - 206I_1^f(0,T)    \right],\\
   \Pi_{T,44}^{hG}(0) &= - g^{\prime 2}\Pi^{SS}_T(m^2_h,m^2_G,T),\\
   \Pi_{L,44}^{hG}(0) &= - g^{\prime 2}\Pi^{SS}_L(m^2_h,m^2_G,T),\\
   \Pi_{T,44}^{GG}(0) &=\frac{g^{\prime 2}}{3}\left[-\frac{1}{2}\left(1+\frac{2}{3}\epsilon\right)I_1^b(m_G^2,T) + m_G^2 \left(1+\frac{2}{3}\epsilon\right)I_2^b(m_G^2,T)  - D_T I_1^b(m_G^2,T)  \right],\\
    \Pi_{L,44}^{GG}(0)   &= -\frac{g^{\prime 2}}{2}\left[I_1^b(m_G^2,T) - 2D_T I_1^b(m_G^2,T) \right]  ,\\
    \Pi_{T,44}^{AH}(0) &= - g^{\prime 2}\Pi^{SS}_T(m^2_A,m^2_H,T),\\
   \Pi_{L,44}^{AH}(0) &= - g^{\prime 2}\Pi^{SS}_L(m^2_A,m^2_H,T),\\
   \Pi_{T,44}^{H^+H^-}(0) &=\frac{g^{\prime 2}}{3}\left[-\frac{1}{2}\left(1+\frac{2}{3}\epsilon\right)I_1^b(m_{H^\pm}^2,T) + m_{H^\pm}^2 \left(1+\frac{2}{3}\epsilon\right)I_2^b(m_{H^\pm}^2,T)  - D_T I_1^b(m_{H^\pm}^2,T)  \right],\\
    \Pi_{L,44}^{H^+H^-}(0)   &= -\frac{g^{\prime 2}}{2}\left[I_1^b(m_{H^\pm}^2,T) - 2D_T I_1^b(m_{H^\pm}^2,T) \right]  ,\\
    \Pi_{T,44}^{GW}(0) &= -\frac{g^2g^{\prime 2}\phi^2}{2}  \Pi_T^{SV}(m_G^2, m_{W,L}^2, m_{W,T}^2, T, 1, 1),\\
    \Pi_{L,44}^{GW}(0) &= -\frac{g^2g^{\prime 2}\phi^2}{2}  \Pi_L^{SV}(m_G^2, m_{W,L}^2, m_{W,T}^2, T), \\
\Pi_{T,44}^{hZ}(0) &= -\frac{g^{\prime 2}\phi^2}{4}  \Pi_T^{SV}(m_h^2, m_{Z,L}^2, m_{Z,T}^2, T, g^\prime s_{w,L} + gc_{w,L}, g^\prime s_{w,T} + gc_{w,T}),\\
    \Pi_{L,44}^{hZ}(0) &= -\frac{g^{\prime 2}(g^\prime s_{w,L} + gc_{w,L})^2\phi^2}{4}  \Pi_L^{SV}(m_h^2, m_{Z,L}^2, m_{Z,T}^2, T),\\
    \Pi_{T,44}^{h\gamma}(0) &= -\frac{g^{\prime 2}\phi^2}{4}  \Pi_T^{SV}(m_h^2, m_{\gamma,L}^2, m_{\gamma,T}^2, T, g^\prime c_{w,L} - gs_{w,L}, g^\prime c_{w,T} - gs_{w,T}),\\
    \Pi_{L,44}^{h\gamma}(0) &= -\frac{g^{\prime 2}(g^\prime c_{w,L} - gs_{w,L})^2\phi^2}{4}  \Pi_L^{SV}(m_h^2, m_{\gamma,L}^2, m_{\gamma,T}^2, T).
\end{align}
The momentum-independent contributions to the (34) component \(\Pi_T\) and \(\Pi_L\) are given by
\begin{align}
    \Pi_{T,34}^{h}&= \Pi_{L34}^{h}= -\frac{gg^\prime}{4}I_1^b(m_h^2,T),\\
       \Pi_{T,34}^G&= \Pi_{L34}^G =  \frac{gg^{\prime }}{4}I^b_1(m_G^2,T),\\
      \Pi_{T,34}^A&= \Pi_{L34}^A =  -\frac{gg^{\prime }}{4}I^b_1(m_A^2,T),\\
       \Pi_{T,34}^H&= \Pi_{L34}^H = - \frac{gg^{\prime }}{4}I^b_1(m_H^2,T),\\
        \Pi_{T,34}^{H^\pm}&=\Pi_{L34}^{H^\pm} =  \frac{gg^{\prime }}{2}I^b_1(m_{H^\pm}^2,T).
\end{align}
The momentum-dependent contributions to the (34) component of \(\Pi_T\) and \(\Pi_L\) are given by
\begin{align}
     \Pi_{T,34}^{ff}(0) &= \frac{1}{4(2-2\epsilon)}gg^{\prime }\sint_k \frac{T_{\mu\nu}\mathrm{tr}[\gamma^\mu (P_L + 4P_R)(\slashed{k}+m_t)\gamma^\nu P_L(\slashed{k}+m_t)]}{(k^2-m_t^2)^2}\nonumber\\
    &\ \ \ + \frac{1}{2(2-2\epsilon)}gg^{\prime }\sint_k \frac{T_{\mu\nu}\mathrm{tr}[\gamma^\mu (P_L + 4P_R)\slashed{k}\gamma^\nu P_L\slashed{k}]}{k^4}\nonumber\\
    &\ \ \ - \frac{3}{4(2-2\epsilon)}gg^{\prime }\sint_k \frac{T_{\mu\nu}\mathrm{tr}[\gamma^\mu (P_L - 2P_R)\slashed{k}\gamma^\nu P_L\slashed{k}]}{k^4}\nonumber\\
    &\ \ \ + \frac{3}{4(2-2\epsilon)}gg^{\prime }\sint_k \frac{T_{\mu\nu}\mathrm{tr}[\gamma^\mu (P_L + 2P_R)\slashed{k}\gamma^\nu P_L\slashed{k}]}{k^4}\nonumber\\
     &\ \ \ - \frac{3}{4(2-2\epsilon)}gg^{\prime }\sint_k \frac{T_{\mu\nu}\mathrm{tr}[\gamma^\mu P_L \slashed{k}\gamma^\nu P_L\slashed{k}]}{k^4}\nonumber\\
    &=  \frac{gg^{\prime}}{6}\left\{\left(1+\frac{1}{6}\right) \left[2D_T I_1^f(m_t^2,T)-I_1^f(m_t^2,T)   \right] - 3(4-2\epsilon)m_t^2I_2^f(m_t^2,T) \right.\nonumber\\
     &\ \ \  \left.    - \left(1-\frac{11}{6}\right)  \left[I_1^f(m_t^2,T) - m_t^2 I_2^f(m_t^2,T)\right] \right\},\\
     \Pi_{L,34}^{ff}(0) &=\frac{gg^\prime}{4}\left[ - 2D_TI_1^f(m_t^2,T)   + 2I_1^f(0,T)    - 3(2-\epsilon)m_t^2 I_2^f(m_t^2,T)\right],\\
     \Pi_{T,34}^{hG}(0) &= gg^{\prime }\Pi^{SS}_T(m^2_h,m^2_G,T),\\
   \Pi_{L,34}^{hG}(0) &= g g^{\prime }\Pi^{SS}_L(m^2_h,m^2_G,T),\\
   \Pi_{T,34}^{GG}(0) &=\frac{gg^{\prime }}{3}\left[-\frac{1}{2}\left(1+\frac{2}{3}\epsilon\right)I_1^b(m_G^2,T) + m_G^2 \left(1+\frac{2}{3}\epsilon\right)I_2^b(m_G^2,T)  - D_T I_1^b(m_G^2,T)  \right],\\
    \Pi_{L,34}^{GG}(0)   &= -\frac{gg^{\prime }}{2}\left[I_1^b(m_G^2,T) - 2D_T I_1^b(m_G^2,T) \right]  ,\\
    \Pi_{T,34}^{AH}(0) &=  gg^{\prime }\Pi^{SS}_T(m^2_A,m^2_H,T),\\
   \Pi_{L,34}^{AH}(0) &= g g^{\prime }\Pi^{SS}_L(m^2_A,m^2_H,T),\\
   \Pi_{T,34}^{H^+H^-}(0) &=\frac{gg^{\prime }}{3}\left[-\frac{1}{2}\left(1+\frac{2}{3}\epsilon\right)I_1^b(m_{H^\pm}^2,T) + m_{H^\pm}^2 \left(1+\frac{2}{3}\epsilon\right)I_2^b(m_{H^\pm}^2,T)  - D_T I_1^b(m_{H^\pm}^2,T)  \right],\\
    \Pi_{L,34}^{H^+H^-}(0)   &= -\frac{gg^{\prime }}{2}\left[I_1^b(m_{H^\pm}^2,T) - 2D_T I_1^b(m_{H^\pm}^2,T) \right]  ,\\
    \Pi_{T,34}^{hZ}(0) &= \frac{gg^{\prime }\phi^2}{4}  \Pi_T^{SV}(m_h^2, m_{Z,L}^2, m_{Z,T}^2, T, g_{Z,L}, g_{Z,T}),\\
    \Pi_{L,34}^{hZ}(0) &= \frac{gg^{\prime }g_{Z,L}^2\phi^2}{4}  \Pi_L^{SV}(m_h^2, m_{Z,L}^2, m_{Z,T}^2, T),\\
    \Pi_{T,34}^{h\gamma}(0) &= \frac{gg^{\prime }\phi^2}{4}  \Pi_T^{SV}(m_h^2, m_{\gamma,L}^2, m_{\gamma,T}^2, T,  g_{\gamma,L}, g_{\gamma,T}),\\
    \Pi_{L,34}^{h\gamma}(0) &= \frac{gg^{\prime }g_{\gamma,L}^2\phi^2}{4}  \Pi_L^{SV}(m_h^2, m_{\gamma,L}^2, m_{\gamma,T}^2, T),
\end{align}
where 
\begin{align}
    g_{Z,L}^2 = (g^\prime s_{w,L} + g c_{w,L}) (g c_{w,L} + g^\prime s_{w,L}),\quad g_{Z,T}^2 = (g^\prime s_{w,T} + g c_{w,T}) (g c_{w,T} + g^\prime s_{w,T}),\\
    g_{\gamma,L}^2 = (g^\prime c_{w,L} - g s_{w,L}) (g s_{w,L} - g^\prime c_{w,L}),\quad g_{\gamma,T}^2 = (g^\prime c_{w,T} - g s_{w,T}) (g s_{w,T} - g^\prime c_{w,T}).
\end{align}
It is straightforward to verify that all components satisfy \(\Pi_L = \Pi_T\) at zero temperature.
    
\end{itemize}
}